\documentclass[review]{elsarticle}
\usepackage[utf8]{inputenc}
\usepackage{hyperref}
\usepackage{lineno}
\usepackage{amsmath}
\usepackage{fixmath}
\usepackage{amsfonts}
\usepackage{amssymb}
\usepackage{array}
\usepackage{float}
\usepackage{graphicx}
\usepackage{subcaption}
\usepackage{booktabs}
\usepackage[a4paper, total={6.6in, 10.4in}]{geometry}
\usepackage{lineno,hyperref}
\usepackage{multirow}
\usepackage{xcolor}
\usepackage{color}
\usepackage{appendix}
\usepackage{gensymb}
\usepackage{makecell}
\usepackage{tikz}
\modulolinenumbers[1]

\begin{document}

\begin{frontmatter}

\title{Cavitation Acoustic Perturbation Equations: A Computational Framework for Source-Resolved Multiphase Hydroacoustics}


\author[mymainaddress]{Zhi Cheng\corref{mycorrespondingauthor}}
\cortext[mycorrespondingauthor]{Corresponding author}
\ead{vamoschengzhi@gmail.com}

\author[mymainaddress]{Rajeev K. Jaiman}


\address[mymainaddress]{Computational Multiphysics Lab, Mechanical Engineering, University of British Columbia, 2054-6250 Applied Science Lane, Vancouver, British Columbia, V6T 1Z4, CANADA.}

\begin{abstract}
\textcolor{black}{
This work develops a cavitation-consistent acoustic perturbation framework for predicting sound generation and propagation in cavitating flows. 
Unlike traditional acoustic perturbation equations developed primarily for single-phase or weakly compressible flows, the proposed formulation embeds cavitation physics directly into the acoustic equation.
By constructing the cavitation acoustic perturbation equations (CAPE), the effects of vapor mass transfer, mixture compressibility, and pressure-rate source are incorporated into a unified acoustic formulation, enabling cavitation-induced noise sources to be directly resolved within the computational domain. The numerical framework is verified through one-dimensional wave-propagation problems, where the solutions become insensitive to further mesh and time-step refinement, and the perfectly matched layer effectively suppresses boundary reflections. The predicted attenuation of acoustic waves over a range of source frequencies follows Stokes’ sound attenuation law, demonstrating the ability of the solver to reproduce physically correct acoustic dissipation. The framework is subsequently applied to cavitating flow past a circular cylinder and a NACA hydrofoil. 
The non-cavitating benchmark exhibits dipole-like radiation associated with unsteady loading, whereas the cavitating configurations produce monopole-like or geometry-modulated radiation associated with volumetric phase-change effects.
Source-term analyses identify the dominant tonal frequencies associated with vortex shedding, cavity shedding, and collapse-induced acoustic excitation. 
The source-term and acoustic-field analyses further show that the phase-change terms introduced in the formulation provide a direct volumetric contribution to the monopole-like source, while localized collapse events are represented through pressure-rate source amplification.
These results demonstrate the capability of the formulation and solver to resolve cavitation-induced tonal singing components.
The proposed framework extends acoustic perturbation methods to cavitating multiphase systems and provides an efficient computational tool for hydroacoustic prediction, source localization, and mechanism analysis in marine and hydraulic applications.
}

\end{abstract}


\end{frontmatter}

\section{Introduction}





Hydrodynamic noise generated by turbulent and cavitating flows remains a critical challenge in marine, hydraulic, and energy systems, with direct implications for structural integrity, environmental impact, and acoustic stealth \citep{blake2017mechanics,brennen2014cavitation,leighton2012acoustic,arndt1981cavitation}. In marine propulsion, cavitation-induced noise is a dominant contributor to underwater radiated sound, particularly under off-design or high-load operating conditions. Among the various cavitation-related noise phenomena, `singing' is especially problematic, manifesting itself as a strong narrow-band tonal noise that can propagate over long distances and lead to unacceptable acoustic signatures \citep{Xincheng2023, Arndt2015, Maines1997, Higuchi1989, brennen2014cavitation,kerr1940problems}. Despite decades of experimental and numerical investigations, the physical origin of singing and its mathematical representation within a unified acoustic framework remain incompletely understood.

Cavitation noise arises from complex, multiscale interactions involving phase change, turbulent flow structures, and vortex behaviors. The formation, shedding, and collapse of vapor cavities introduce highly localized pressure transients and volumetric fluctuations, which act as efficient acoustic sources. Unlike non-cavitating flow noise, which is often dominated by dipole sources associated with unsteady surface loading, cavitating flows can generate strong monopole-like radiation due to rapid volume change \citep{brennen2014cavitation,brennen2005fundamentals}. These features make cavitation noise fundamentally different in nature and considerably more challenging to model and predict. In practical marine propellers, the situation is further complicated by blade rotation, wake non-uniformity, and strong coupling between hydrodynamic, cavitation, and acoustic processes, all of which can contribute to the emergence of tonal singing \citep{Ross1989,carlton2018marine}.

From a numerical standpoint, several approaches have been developed to predict flow-induced noise. Direct noise computation, in which the compressible Navier-Stokes equations are solved to resolve both flow and acoustic scales, provides the most complete description of the physics \citep{seo2008prediction,colonius2004computational,freund2001noise}. However, this approach is prohibitively expensive for realistic, low-Mach-number flows, especially when cavitation is present, due to the extreme disparity between hydrodynamic and acoustic length and time scales. As a result, direct simulations are typically limited to simplified configurations or short time windows.

Hybrid methods have therefore become the dominant tools for engineering aero- and hydroacoustics. Among these, the Ffowcs Williams-Hawkings (FW-H) acoustic analogy and its variants have been widely used to predict far-field noise based on near-field flow information \citep{ffowcs1969sound,farassat1998acoustic,farassat1975theory,farassat1981linear,farassat1988uses}. While FW-H methods are effective for non-cavitating flows and surface-dominated noise sources, they suffer from inherent limitations when applied to cavitating flows. In particular, the classical FW-H formulation relies on equivalent surface and volume source terms derived under assumptions that do not explicitly account for phase change. As a consequence, it is difficult to accurately represent cavitation-induced monopole sources, especially those arising from localized vapor generation and collapse within the flow field \citep{sezen2023marine}. Although several extensions and empirical corrections have been proposed, the lack of a rigorous treatment of cavitation physics remains a fundamental drawback of FW-H-based approaches for predicting cavitation noise and singing.

An alternative class of methods is based on acoustic perturbation equations (APE) \citep{shen2004collocated,ewert2003acoustic} and closely related linearized perturbed compressible equations (LPCE) \citep{seo2006linearized,seo2007aerodynamic}. These approaches decompose the flow field into hydrodynamic and acoustic components and solve a dedicated acoustic equation driven by source terms extracted from the flow solution. Compared with FW-H methods, APE-based formulations offer several important advantages: they allow acoustic waves to be generated and propagated directly within the computational domain, they can capture volumetric noise sources without relying on surface integrals, and they are well suited for systems where the sound source location cannot be predicted \citep{ewert2003acoustic,zhang2006hybrid,shur2008hybrid}. Over the past two decades, APE and related formulations have been successfully applied to a wide range of aeroacoustic problems, including sound generation in complex geometries and low-speed flows. Representative examples include high-order immersed boundary methods for acoustic wave scattering and low-Mach-number flow-induced sound in complex geometries, as well as collocated-grid finite-volume formulations for aeroacoustic computations of low-speed flows \citep{seo2011high,shen2004collocated}.

Despite these advances, existing APE-based formulations have been almost exclusively developed for single-phase or weakly compressible flows \citep{tam2004computational} and do not explicitly incorporate cavitation effects. In their classical form, the governing equations assume a single-valued equation of state and neglect mass transfer between phases. As a result, the dominant acoustic contributions associated with cavitation---namely rapid vaporization, condensation, and cavity collapse---are not represented in a mathematically consistent manner. 
Hence, the above studies specifically focusing on cavitation noise often rely on idealized assumptions, empirical source models, or fully compressible simulations that remain computationally expensive and difficult to generalize.

\textcolor{black}{
Consequently, a gap still exists between high-fidelity cavitating-flow simulations and computational acoustic models capable of consistently coupling cavitation dynamics with acoustic wave propagation.
Existing FW-H formulations provide an efficient far-field acoustic framework but do not directly resolve cavitation-induced volume noise source processes within the flow domain. Existing APE and LPCE approaches enable in-domain acoustic-source analysis, yet their formulations are generally developed for single-phase or weakly compressible flows and do not explicitly account for cavitation mass transfer and the associated variations in mixture compressibility. As a result, the physical link between cavitation dynamics and acoustic radiation remains insufficiently understood, particularly for singing-like tonal phenomena.
}

The objective of the present work is to establish a cavitation-aware acoustic perturbation framework that connects cavitation dynamics, acoustic generation, and sound propagation within a unified formulation. By extending the conventional APE framework to incorporate phase-change-induced mass transfer and mixture-compressibility effects, a closed acoustic control equation suitable for cavitating flows is derived and implemented within a coupled flow-acoustics solver. 
Beyond far-field prediction, the formulation enables quantitative evaluation of cavitation-induced acoustic source terms and their physical contribution to the perturbation field.
In particular, the framework enables simultaneous examination of cavity evolution, pressure-rate amplification, acoustic source generation, and sound radiation, 
allowing the resulting acoustic source distribution and tonal components to be quantified and interpreted within the same numerical framework.
The main contributions of this work are threefold. First, a cavitation APE (CAPE) formulation is derived for a homogeneous liquid-vapor mixture by introducing the compressibility of the mixture and the phase-change mass transfer into the acoustic perturbation equations. An acoustic energy balance is also obtained to clarify the role of cavitation-induced mass transfer as a volumetric monopole-like acoustic source. Second, the formulation is implemented in a finite-volume predictor-corrector solver with acoustic subcycling and absorbing-layer treatment for long-range wave propagation. Third, the framework is verified using one-dimensional propagation tests and then applied to canonical cavitating cylinder and hydrofoil configurations, where it identifies the cavitation-driven tonal radiation associated with localized cavity collapse.


The remainder of this paper is organized as follows. Section~2 introduces the cavitating base-flow equations, derives the CAPE formulation with PML treatment, and states the one-way coupling and the acoustic-energy relations. Section~3 presents the finite-volume predictor-corrector scheme and the multirate flow-acoustic coupling
procedure. Section~4 first verifies the numerical properties of the solver and subsequently applies the framework to non-cavitating and cavitating cylinder flows and to a cavitating hydrofoil. Section~5 summarizes the main conclusions.

\section{Mathematical Formulations}

\subsection{Cavitation Acoustic Perturbation Equations}
In this section, we derive cavitation acoustic perturbation equations (CAPE) for a homogeneous liquid-vapor mixture. The flow solution obtained from the incompressible cavitating-flow solver is interpreted as an unsteady hydrodynamic base state, while the acoustic field is represented by small compressible perturbations superimposed on this base state.

The phase indicator $\alpha _l$($\boldsymbol{x}$,$t$) represents the phase fraction of the liquid phase in the fluid mixture, where $\boldsymbol{x}$ and $t$ are spatial and temporal coordinates. 
The subscripts `$_l$' and `$_v$' represent the liquid phase and the gas phase, respectively.
The density $\rho_0$ and dynamic viscosity $\mu$ of the fluid are obtained as a linear weighted combination of the liquid and vapor phases:
\begin{linenomath}
	\begin{equation}\label{}
\rho_0 =\rho _{0,l}\alpha _l+\rho _{0,v} \left(1 - \alpha _l \right) ,
	\end{equation}
\end{linenomath}
\begin{linenomath}
	\begin{equation}\label{}
\mu =\mu _l\alpha _l+\mu _v \left(1 - \alpha _l \right) .
	\end{equation}
\end{linenomath}


The mixture density is decomposed into an incompressible base-state component and a compressible perturbation:
\begin{linenomath}
\begin{equation}\label{cb_1}
\rho =\rho _{0} + \rho' = \alpha _l\rho _{0,l}+\left( 1-\alpha _l \right) \rho _{0,v} + \alpha _l\rho'_l+\left( 1-\alpha _l \right) \rho'_v	,
\end{equation}
\end{linenomath}
and the compressible Navier-Stokes equations are written as:
\begin{equation}\label{LES_eq1}
	\frac{\partial \rho}{\partial t}+\frac{\partial \left( \rho u_i \right)}{\partial x_i}=\rho \left( \frac{1}{\rho _l}-\frac{1}{\rho _v} \right) \dot{m},
\end{equation}
\begin{equation}\label{LES_eq2}
\frac{\partial \left( \rho u_i \right)}{\partial t}+\frac{\partial \left( \rho u_i u_j \right)}{\partial x_j}=-\frac{\partial p}{\partial x_j}+\frac{\partial}{\partial x_j}\left( \mu_{\mathrm{eff}} \frac{\partial u_i}{\partial x_j} \right) ,
\end{equation}
velocity and pressure are similarly decomposed as an incompressible part and a compressible perturbation part:
\begin{equation}\label{cb_2}
u_i = U _i + u'_i,
\end{equation}
\begin{equation}\label{cb_3}
p = P + p'.
\end{equation}

In the present hybrid formulation, the cavitating flow solution is interpreted as a slowly varying hydrodynamic base state, denoted by \((\rho_0,\mathbf{U},P,\alpha_l,\dot m)\). Acoustic variables are then introduced as small compressible perturbations \((\rho',\mathbf{u}',p')\) superimposed on this base state. The decomposition is therefore not intended to solve the fully compressible multiphase flow directly; rather, it provides a perturbation system driven by the base-state cavitation dynamics.
Inserting Eqs. \ref{cb_1},\ref{cb_2}, and \ref{cb_3} into Eqs. \ref{LES_eq1},\ref{LES_eq2}:
\begin{equation}\label{cns_1}
\frac{\partial \left( \rho _0+\rho ' \right)}{\partial t}+\frac{\partial \left( \left( \rho _0+\rho ' \right) \left( \text{U}_{\text{i}}+u_i' \right) \right)}{\partial x_i}=\left( \rho _0+\rho ' \right) \left( \frac{1}{\rho _l}-\frac{1}{\rho _v} \right) \dot{m},
\end{equation}
\begin{linenomath}
\begin{align}\label{cns_2}
\frac{\partial \left( \left( \rho _0+\rho ' \right) \left( \text{U}_{\text{i}}+u_i' \right) \right)}{\partial t} + \frac{\partial ((\rho_0 + \rho')(U_i+u'_i)(U_j+u'_j))}{\partial x_j} 
\\= -\frac{\partial (P + p')}{\partial x_j}+\frac{\partial}{\partial x_j}\left( \mu_{\mathrm{eff}} \frac{\partial u_i}{\partial x_j} \right) .
\end{align}
\end{linenomath}

Subtracting the corresponding base-state cavitating flow equations (cf. with compressible counterpart Eqs. \ref{LES_eq1},\ref{LES_eq2}) from Eqs. \ref{cns_1},\ref{cns_2} yields a set of nonlinear equations for the acoustic fluctuations:
\begin{equation}\label{}
f_i=\rho _0u_i'+\rho '\text{U}_{\text{i}}+\rho 'u_i',
\end{equation}
\begin{equation}\label{1212}
	\frac{\partial \rho '}{\partial t}+\frac{\partial f_i}{\partial x_i}=h,
\end{equation}
\begin{equation}\label{hi_1}
h=(\rho_0 + \rho') \left( \frac{1}{\rho _l}-\frac{1}{\rho _v} \right) \dot{m} - (\rho_0) \left( \frac{1}{\rho _{0,l}}-\frac{1}{\rho _{0,v}} \right) \dot{m},
\end{equation}
\begin{equation}\label{}
\frac{\partial f_i}{\partial t}+\frac{\partial}{\partial x_j}\left[ f_i\left( U_j+u'_j \right) +\rho _0U_iu'_j+p'\delta _{ij} \right] - \frac{\partial \tau _{ij}^{\prime}}{\partial x_j} =0,
\end{equation}
where 
\begin{equation}\label{}
\tau _{ij}^{\prime}= \mu_{\mathrm{eff}} \left( \frac{\partial u_{i}^{\prime}}{\partial x_j}+\frac{\partial u_{j}^{\prime}}{\partial x_i}-\frac{2}{3}\delta _{ij}\frac{\partial u_{k}^{\prime}}{\partial x_k} \right).
\end{equation}
$f_i$, $\delta _{ij}$, $\tau _{ij}$ are the acoustic co-velocity components, the Kronecker delta function, and the perturbed viscous stresses, respectively. $h$ represents the source term of mass-transfer induced by cavitation. Using the pressure-density relation adopted in the perturbation
formulation:
\begin{equation}\label{}
\frac{\partial p}{\partial \rho}=c^2,
\end{equation}
\begin{equation}\label{}
\frac{\partial p }{\partial t}=\frac{\partial p }{\partial \rho}\frac{\partial \rho}{\partial t}=c^2\frac{\partial \rho}{\partial t}=c^2\frac{\partial \left( \rho ^ \prime+\rho _0 \right)}{\partial t},
\end{equation}
with $p$ = $ P+p^{\prime} $:
\begin{equation}\label{APE_prho}
\frac{\partial p^{\prime}}{\partial t}-c^2\frac{\partial \rho ^{\prime}}{\partial t}=c^2\frac{\partial \rho _0}{\partial t}-\frac{\partial P}{\partial t}.
\end{equation}
For the two-phase homogeneous mixture, the acoustic closure is introduced through the compressibility of the mixture, rather than through a single-phase ideal-gas relation. Assuming mechanical equilibrium between the liquid and vapor phases, the compressibility of the mixture is written as
\begin{equation}
\label{c_updated}
\frac{1}{\rho_0 c^2}
=
\frac{\alpha_l}{\rho_{0,l} c_l^2}
+
\frac{1-\alpha_l}{\rho_{0,v} c_v^2},
\end{equation}
where \(c_l\) and \(c_v\) are the sound speeds of the pure liquid and vapor phases, respectively. This Wood-type relation shows that the local acoustic speed is strongly controlled by the vapor volume fraction and may decrease significantly in the cavitating region. 

Combining Eqs.~\ref{1212} and \ref{APE_prho} gives:
\begin{equation}\label{}
\frac{\partial p'}{\partial t} + c^2\frac{\partial f_i}{\partial x_i} =c^2\frac{\partial \rho _0}{\partial t} -\frac{\partial P}{\partial t} + c^2 h.
\end{equation}
The cavitation-induced mass-transfer source in Eq.~\ref{hi_1} can be rewritten as
\begin{equation}\label{hi_2}
h=\left( \rho _0+\rho ' \right) \left( \frac{\rho _v-\rho _l}{\rho _l\rho _v} \right) \dot{m}-\left( \rho _0 \right) \left( \frac{\rho _{0,v}-\rho _{0,l}}{\rho _{0,l}\rho _{0,v}} \right) \dot{m},
\end{equation}
where $\dot{m} = \dot{m}_v+\dot{m}_c$, with
\begin{linenomath}
\begin{equation}\label{eqm1}
\dot{m}_c=C_c\alpha _l\left( 1-\alpha _l \right) \frac{3\rho _v\rho _l}{\rho R_B}\sqrt{\frac{2}{3\rho _l\lvert p-p_v \rvert}}\text{max}\left( p-p_v,0 \right),
\end{equation}
\end{linenomath}

\begin{linenomath}
\begin{equation}\label{eqm2}
\dot{m}_v=C_v\alpha _l\left( 1 + \alpha_{Nuc} -\alpha _l \right) \frac{3\rho _v\rho _l}{\rho R_B}\sqrt{\frac{2}{3\rho _l \lvert p-p_v \rvert}} \text{min}\left( p-p_v,0 \right),
\end{equation}
\end{linenomath}
where $C_c$ and $C_v$ are the coefficients of the condensation and vaporization models, respectively, and $p_v$ is the fluid vapor pressure under saturation conditions. $\alpha_{Nuc}$ and $R_B$ are the volume fraction and radius of the bubble nuclei in the fluid and are given as
\begin{linenomath}
\begin{equation}\label{alphaNuc}
\alpha _{Nuc}=\left( n_0\pi\frac{d_{Nuc}^3}{6}  \right) /\left( 1+n_0\pi\frac{d_{Nuc}^3}{6} \right) ,
\end{equation}
\end{linenomath}

\begin{linenomath}
\begin{equation}\label{Rb}
R_B=\sqrt[3]{\frac{3}{4 \pi n_0} \frac{1+\alpha_{Nuc}-\alpha_l}{\alpha_l}} ,
\end{equation}
\end{linenomath}
where $n_0$ is the number of nuclei per unit volume and $d_{Nuc}$ is the corresponding diameter of the nuclei.
In the present simulations, the Schnerr-Sauer model constants are set to \(C_c=1.0\) and \(C_v=1.0\). The density of nuclei number is prescribed as
\(n_0=1.6\times 10^{13}\,\mathrm{m}^{-3}\), and the diameter of the nuclei is set to
\(d_{\mathrm{Nuc}}=6\times 10^{-5}\,\mathrm{m}\). The volume fraction of the corresponding nuclei
 \(\alpha_{\mathrm{Nuc}}\) is calculated from Eq.~\ref{alphaNuc} and is therefore not treated as an independent model parameter.
The sign convention is chosen so that positive $\dot{m}$ corresponds to the net contribution of the condensation/source to the mixture continuity equation, while negative $\dot{m}$ corresponds to the vaporization; this convention is used consistently in the CAPE source term $h$.

The final CAPE can be written as
\begin{equation}\label{}
f_i=\rho _0u_i'+\rho '\text{U}_{\text{i}}+\rho 'u_i', \,\, h=\left( \rho _0+\rho ' \right) \left( \frac{\rho _v-\rho _l}{\rho _l\rho _v} \right) \dot{m}-\left( \rho _0 \right) \left( \frac{\rho _{0,v}-\rho _{0,l}}{\rho _{0,l}\rho _{0,v}} \right) \dot{m},
\end{equation}
\begin{equation}\label{}
	\frac{\partial \rho '}{\partial t}+\frac{\partial f_i}{\partial x_i}=h,
\end{equation}
\begin{equation}\label{f_func0}
\frac{\partial f_i}{\partial t}+\frac{\partial}{\partial x_j}\left[ f_i\left( U_j+u'_j \right) +\rho _0U_iu'_j+p'\delta _{ij} \right]  - \frac{\partial \tau _{ij}^{\prime}}{\partial x_j}=0,
\end{equation}
\begin{equation}\label{finalAPE_P}
\frac{\partial p'}{\partial t} - c^2\frac{\partial \rho'}{\partial t} = c^2\frac{\partial \rho _0}{\partial t}-\frac{\partial P}{\partial t}.
\end{equation}

\subsection{Non-Reflecting Method: Perfectly Matched Layer}
Because the present APE system resolves acoustic-wave propagation directly in the computational domain, the treatment of outgoing waves at the outer boundary is critical. Standard non-reflecting boundary conditions may still generate residual reflections when applied to broadband or weakly damped acoustic fields. Therefore, the present work employs a PML-type absorbing layer surrounding the physical acoustic domain. In this layer, damping terms are added to the perturbation variables so that outgoing waves are attenuated before reaching the outer boundary, where a conventional non-reflecting condition is imposed.
The equation of standard non-reflecting boundary conditions \cite{seo2011high} for the subsonic situation is given as:
\begin{linenomath}
\begin{equation}\label{eq:ns1}
\frac{\partial \varphi}{\partial t}+U_n\frac{\partial \varphi}{\partial n}=0,
\end{equation}
\end{linenomath}
\begin{linenomath}
\begin{equation}\label{eq:ns2}
U_n=u_n-c,
\end{equation}
\end{linenomath}
with \(\varphi\) as the parameters treated by boundary conditions, \(c\) as local sound velocity, \(u_n\) as velocity normal to the boundary.
Based on our numerical observations, traditional Non-Reflecting Boundary Conditions still produce little reflection from the boundary. Therefore, this study will use the Perfectly Matched Layer (PML) method plus the traditional Non-Reflecting Boundary Condition. The PML method defines an attenuation region surrounding the original computational domain, and adds the damping terms into the CAPE equations:
\begin{equation}\label{final_c_eq_0}
f_i=\rho _0u_i'+\rho ' U_{\text{i}}+\rho 'u_i', \,\, h=\left( \rho _0+\rho ' \right) \left( \frac{\rho _v-\rho _l}{\rho _l\rho _v} \right) \dot{m}-\left( \rho _0 \right) \left( \frac{\rho _{0,v}-\rho _{0,l}}{\rho _{0,l}\rho _{0,v}} \right) \dot{m},
\end{equation}
\begin{equation}\label{final_c_eq_1}
	\frac{\partial \rho '}{\partial t} +c \sigma \rho '+\frac{\partial f_i}{\partial x_i}=h,
\end{equation}
\begin{equation}\label{final_c_eq_2}
\frac{\partial f_i}{\partial t}+c \sigma f_i +\frac{\partial}{\partial x_j}\left[ f_i\left( U_j+u'_j \right) +\rho _0U_iu'_j+p'\delta _{ij} \right]  - \frac{\partial \tau _{ij}^{\prime}}{\partial x_j}=0,
\end{equation}
\begin{equation}\label{final_c_eq}
\frac{\partial p'}{\partial t}+c \sigma p' - c^2\frac{\partial \rho'}{\partial t} = c^2\frac{\partial \rho _0}{\partial t}-\frac{\partial P}{\partial t},
\end{equation}
where $\sigma$ is the damping coefficient. The damping coefficient $\sigma$ gradually increases from zero at the
interface between the effective acoustic region and the PML to a prescribed maximum value $\sigma_{\max}$ at the outer limit.

\subsection{Governing Equations of Incompressible Cavitating Flow}

The incompressible cavitating flow field is obtained using an unsteady Reynolds-averaged Navier-Stokes (URANS) formulation coupled with the CAPE solver. The objective of the present study is to develop and verify the acoustic perturbation formulation for cavitating flows and to demonstrate its ability to resolve cavitation-induced source terms. Since the dominant acoustic forcing considered in the two-dimensional case here is associated with phase-change mass transfer, mixture-compressibility variation, vortex-shedding, and pressure-rate amplification, a URANS-based base flow provides a practical and consistent input for the present hybrid framework. Similar RANS/APE strategies have been widely used in aeroacoustic computations of non-cavitating flows \citep{tucker2013unsteady,rosa2017comparison,ewert2011caa,engel2014application,TrailingedgenoiseAPE}, including trailing-edge noise, jet noise, and broadband source modeling. Extension to three-dimensional turbulence-resolving base flows is left for future work.

The incompressible cavitating flow is modeled using the RANS equations with cavitation-induced mass transfer. The continuity equation is written as follows:
\begin{equation}
\frac{\partial U_j}{\partial x_j}
=
\left(
\frac{1}{\rho_{0,l}} - \frac{1}{\rho_{0,v}}
\right)\dot{m},
\label{eq:continuity}
\end{equation}
where $\dot{m}$ denotes the net mass transfer rate due to cavitation.
\(x_i\) is the $i$-th component of a Cartesian coordinate vector $\boldsymbol{x}$ with $i$, $j \equiv$ 1, 2, 3 corresponding to the $x$-, $y$- and $z$-directions, respectively; and \(u_i\) represents the $i$-th component of fluid velocity.
The RANS momentum equations take the form of
\begin{equation}
\frac{\partial (\rho_0 U_i)}{\partial t}
+
\frac{\partial (\rho_0 U_i U_j)}{\partial x_j}
=
-
\frac{\partial P}{\partial x_j}
+
\frac{\partial}{\partial x_j}
\left(
\mu_{\mathrm{eff}}
\frac{\partial U_i}{\partial x_j}
\right),
\label{eq:momentum}
\end{equation}
where $\mu_{\mathrm{eff}}=\mu+\mu_t$ is the effective viscosity, consisting of the molecular viscosity $\mu$ and the eddy viscosity $\mu_t$ provided by the turbulence model.
The cavitation mass transfer rate is decomposed as
\begin{equation}
\dot{m} = \dot{m}_c + \dot{m}_v ,
\label{eq:masstransfer}
\end{equation}
where $\dot{m}_c$ and $\dot{m}_v$ denote the condensation and vaporization rates, respectively.

Turbulence effects are modeled using a RANS-based closure, with the eddy viscosity $\mu_t$ obtained from the $k$-$\omega$ SST model \citep{Menter1994}. This model is selected for its robustness in adverse pressure-gradient flows and near-wall regions, which are critical for cavitation inception, shedding, and collapse. 
The vaporization and condensation rates, $\dot{m}_v$ and
$\dot{m}_c$, are evaluated using the Schnerr-Sauer model \citep{sauer2001development}, which is
derived from a simplified Rayleigh-Plesset equation and has been widely applied to cavitating flows \citep{asnaghi2017improvement, kashyap2023unsteady}.


\subsection{One-Way Flow-Acoustic Coupling Strategy}

Equations \ref{final_c_eq_0} -\ref{final_c_eq} show that the timewise variation of incompressible pressure $P$, mass transfer $\dot{m}$, and velocity $U_i$ acts as sources for the generation and propagation of acoustic perturbation ($p'$, $ \rho '$, and $u_i'$) respectively. However, the solutions for $P$, $\dot{m}$, and $U_i$ (Eqs. \ref{eq:continuity} and \ref{eq:momentum}) are independent
of $p'$, $ \rho '$, and $u_i'$. Thus, the above mathematical formulation involves one-way coupling between flow and acoustics. 
The assumption of one-way coupling is justified by the smallness of acoustic feedback relative to flow pressure and momentum scales \cite{ewert2003acoustic,shen2004collocated}. 
In the present cases, the acoustic pressure amplitude satisfies
\begin{equation}
\epsilon_p=\frac{|p'|_{\mathrm{rms}}}{\rho_l U_0^2}\ll 1.
\end{equation}

The acoustic velocity perturbation remains much smaller than the hydrodynamic velocity scale. Therefore, the feedback of \((p',\rho',u_i')\) on the cavitating flow solution is of higher order and is ignored. The acoustic field is thus driven by base-state quantities \((P,U_i,\alpha_l,\dot m)\), while acoustic perturbations do not modify the cavitation dynamics.
If an unsteady flow
reaches a periodic state, the periodic flow solution is first obtained, and thereafter the acoustic simulation is initiated.

\subsection{Acoustic Energy Balance and Monopole Source Interpretation}

In order to further clarify the physical role of cavitation-induced mass transfer in the present cavitation-aware acoustic perturbation formulation, an acoustic energy balance is derived in this section.
Because the mixture density $\rho_0$ and sound speed $c$ vary in
cavitating regions, the following energy balance is derived under a
locally frozen-coefficient approximation. Specifically, $\rho_0$ and
$c$ are evaluated at their instantaneous local values but are treated
as locally constant when forming the quadratic acoustic energy density and flux. The resulting relation is therefore interpreted as a local acoustic-energy diagnostic rather than an exact conservation law for the fully variable-coefficient system.
For clarity, a simplified form of the governing equations is considered by neglecting viscous dissipation and mean-flow convection. The CAPE equations are then written as:

\begin{equation}
\frac{\partial \rho ^ \prime}{\partial t}+\nabla \cdot \boldsymbol{f}=h,
\end{equation}

\begin{equation}
\frac{\partial \boldsymbol{f}}{\partial t}+\nabla p ^ \prime=0,
\end{equation}

\begin{equation}
\frac{\partial p'}{\partial t}
+
c^2 \nabla \cdot \boldsymbol{f}
=
-
\frac{\partial P}{\partial t}
+
c^2 h
+
c^2
\frac{\partial \rho_0}{\partial t}.
\end{equation}

Multiplying the momentum equation by \(\boldsymbol{f}/\rho_0\) and the pressure equation by \(p'/(\rho_0 c^2)\), and subsequently combining the resulting expressions, yields the following acoustic energy equation:

\begin{equation}
\frac{\partial}{\partial t}
\left(
\frac{p'^2}{2\rho_0 c^2}
+
\frac{\boldsymbol{f}\cdot\boldsymbol{f}}{2\rho_0}
\right)
+
\nabla \cdot
\left(
\frac{p'\boldsymbol{f}}{\rho_0}
\right)
=
\frac{p'}{\rho_0}
\left(
h
+
\frac{\partial \rho_0}{\partial t}
-
\frac{1}{c^2}
\frac{\partial P}{\partial t}
\right).
\end{equation}

The left-hand side consists of the temporal variation of the acoustic energy density and the divergence of the acoustic energy flux, respectively. The first term inside the brackets corresponds to compressible acoustic potential energy, while the second term represents the acoustic kinetic energy associated with the field \(\boldsymbol{f}\). The right-hand side represents the volumetric acoustic power input generated by cavitation dynamics and compressibility variations of the background-flow. In particular, the term $\frac{p' h}{\rho_0}$
describes the injection of acoustic energy associated with phase-change-induced volumetric variation. Since this contribution enters directly through the continuity and compressibility relations rather than through fluctuating surface forces, it behaves as a monopole-like volumetric acoustic source in the sense of Lighthill's acoustic analogy. Physically, rapid vapor generation and collapse locally modify the volume and compressibility of the mixture, producing strong pressure transients that radiate acoustic waves into the surrounding medium.
 This energy balance therefore provides a direct mechanism-level interpretation of the CAPE source terms and motivates the finite-volume implementation described next.

\section{Numerical Discretization and Solution Procedure}
The incompressible cavitating-flow equations are discretized using a
cell-centered finite-volume method and advanced using the transient PIMPLE algorithm, which combines the semi-implicit method for pressure-linked equations (SIMPLE) \cite{PATANKAR19721787} and the pressure-implicit with the splitting of operators (PISO) algorithm \cite{issa1986solution}. All of these algorithms are iterative solvers, but PISO and PIMPLE are used for transient problems, whereas SIMPLE is used for steady-state problems. The pressure-velocity coupling provided by the PIMPLE algorithm improves robustness.
The CAPE system is solved using a segregated finite-volume predictor-corrector
procedure. This section presents the cell-centered algebraic system,
the pressure-correction procedure, and the multirate coupling between
the flow and acoustic solvers. The corresponding control-volume
integration, face interpolation, and face-level correction are provided
in \ref{app:face_discretization}.
For clarity, the following discretization is written for the effective acoustic region, where $\sigma=0$. Within the PML region, the damping terms in Eqs.~\ref{final_c_eq_1}-\ref{final_c_eq} are assembled into the corresponding diagonal
and source contributions without altering the predictor-corrector
structure.

\subsection{Finite-Volume Predictor-Corrector Scheme}
\label{sec_PC}
The acoustic co-velocity equation, Eq.~\ref{f_func0}, is written in vector form as
\begin{equation}
\frac{\partial \boldsymbol{f}}{\partial t}
+
\nabla\cdot
\left[
\boldsymbol{f}
\left(
\boldsymbol{U}+\boldsymbol{u}'
\right)
+
\rho_0\boldsymbol{U}\boldsymbol{u}'
\right]
=
-\nabla p'
+
\nabla\cdot\boldsymbol{\tau}'.
\label{eq:cape_momentum_vector}
\end{equation}

Let $t$ denote the current acoustic time level and
$t+\Delta t$ the next time level. The nonlinear terms containing
$\boldsymbol{u}'$ are evaluated using the latest available acoustic
fields, while the predicted co-velocity $\boldsymbol{f}^{*}$ remains
implicit. After control-volume integration and face interpolation, the
discretized predictor equation for cell $P$ is written as
\begin{equation}
A_P^{t}\boldsymbol{f}_P^{*}
+
\sum_{N\in\mathcal{N}(P)}
A_N^{t}\boldsymbol{f}_N^{*}
-
\boldsymbol{E}_P^{t}
=
-\left(\nabla p'^{\,t}\right)_P ,
\label{eq:co_velocity_predictor}
\end{equation}
where $A_P^{t}$ and $A_N^{t}$ denote the diagonal and neighboring-cell
coefficients, respectively, and $\boldsymbol{E}_P^{t}$ contains the
remaining explicitly evaluated terms. The detailed expressions leading
to Eq.~\eqref{eq:co_velocity_predictor} are given in~\ref{app:face_discretization}.

The non-pressure contribution to the predicted co-velocity is defined as
\begin{equation}
\mathrm{\mathcal{H}}_P^{*}
=
\frac{1}{A_P^{t}}
\left(
-\sum_{N\in\mathcal{N}(P)}
A_N^{t}\boldsymbol{f}_N^{*}
+
\boldsymbol{E}_P^{t}
\right).
\label{eq:hbyaf}
\end{equation}
The predicted field does not generally satisfy the discrete acoustic
continuity constraint,
\begin{equation}
\nabla\cdot\boldsymbol{f}
+
\frac{1}{c^2}
\frac{\partial p'}{\partial t}
=
S_N,
\label{eq:acoustic_continuity_constraint}
\end{equation}
where the combined CAPE source term is
\begin{equation}
S_N
=
h
+
\frac{\partial \rho_0}{\partial t}
-
\frac{1}{c^2}
\frac{\partial P}{\partial t}.
\label{eq:combined_cape_source}
\end{equation}
At the $o$-th acoustic substep, the corrected pressure
$p'^{\,*}$ is obtained by solving
\begin{equation}
-\nabla\cdot
\left(
\frac{1}{A^{t}}
\nabla p'^{\,*}
\right)
+
\frac{1}{c^2}
\left(
\frac{p'^{\,*}-p'^{\,t}}{\Delta t}
\right)
=
-\nabla\cdot\mathrm{\mathcal{H}}^{*}
+
S_N^{\,t+o\Delta t}.
\label{eq:acoustic_pressure_correction}
\end{equation}
The co-velocity is subsequently corrected according to
\begin{equation}
\boldsymbol{f}_P^{**}
=
\mathrm{\mathcal{H}}_P^{*}
-
\frac{1}{A_P^{t}}
\left(\nabla p'^{\,*}\right)_P .
\label{eq:co_velocity_correction}
\end{equation}

The pressure and co-velocity corrections are repeated until the
prescribed residual tolerances are satisfied. In the present
simulations, two to four correction iterations are generally sufficient
to obtain $\boldsymbol{f}^{\,t+\Delta t}$ and
$p'^{\,t+\Delta t}$.
After convergence of the pressure-correction loop, the acoustic density
perturbation is updated as
\begin{equation}
\rho'^{\,t+\Delta t}
=
\rho'^{\,t}
+
\frac{
p'^{\,t+\Delta t}
+
P^{\,t+\Delta t}
-
p'^{\,t}
-
P^{\,t}
}{
1.5c^{2,t}
-
0.5c^{2,t-\Delta t}
}
-
\left(
\rho_0^{\,t+\Delta t}
-
\rho_0^{\,t}
\right).
\label{eq:density_update}
\end{equation}
The acoustic velocity is then recovered from the definition of the
co-velocity:
\begin{equation}
\boldsymbol{u}'^{\,t+\Delta t}
=
\frac{
\boldsymbol{f}^{\,t+\Delta t}
-
\rho'^{\,t+\Delta t}
\boldsymbol{U}^{\,t+\Delta t}
}{
\rho_0^{\,t+\Delta t}
+
\rho'^{\,t+\Delta t}
}.
\label{eq:acoustic_velocity_update}
\end{equation}
The local mixture sound speed is subsequently updated using Eq.~\ref{c_updated}.

\subsection{Multirate Subcycling for Flow-Acoustic Coupling}
The characteristic time scale of the acoustic field is smaller than
that of the hydrodynamic field. The cavitating-flow solver therefore
uses a time step $\Delta t_f$, while the CAPE solver is advanced using
the acoustic time step
\begin{equation}
\Delta t
=
\frac{\Delta t_f}{m},
\label{eq:acoustic_substep}
\end{equation}
where $m$ is the number of acoustic substeps within one hydrodynamic
time step.

\begin{figure}
	\centering
	\centering\includegraphics[width=0.6\linewidth]{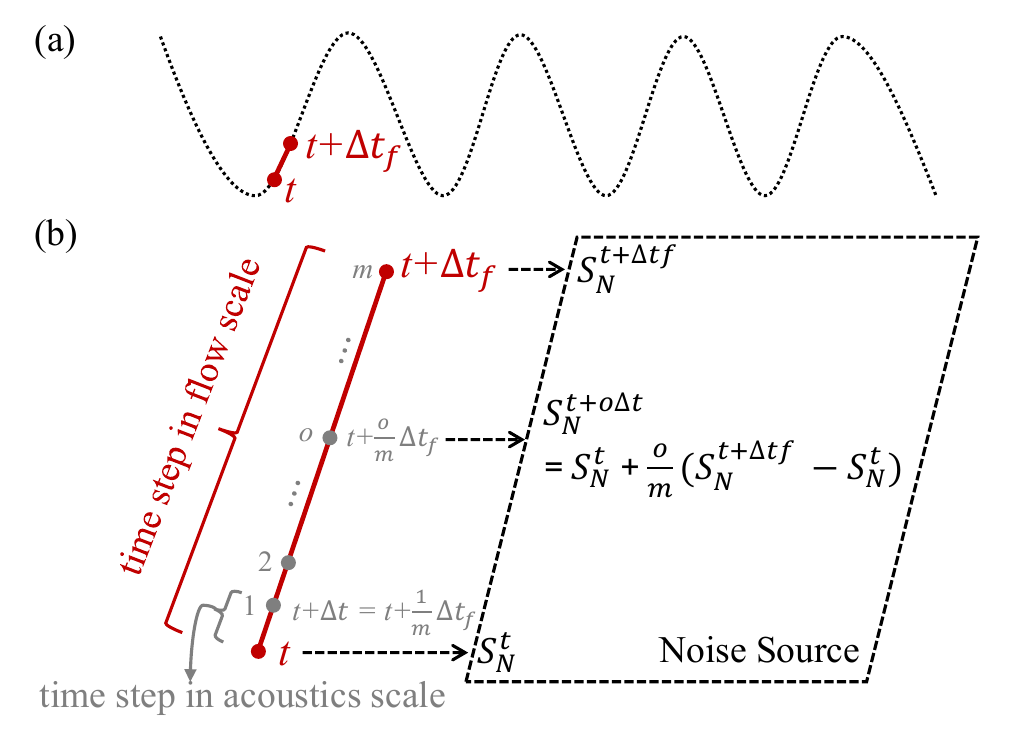}
	\caption{(a) A specific time step from $t$ to $t + \Delta t_f$ in the periodic variation of the flow field. (b) Enlarged view of the chosen flow time step $\Delta t_f$, with acoustics time step $\Delta t$ (= $\Delta t_f /m$) and noise sources at corresponding time steps.}
	\label{diff_timestep_diagram}
\end{figure}

As illustrated in Fig.~\ref{diff_timestep_diagram}, the CAPE source term is available from the flow solution at $t$ and $t+\Delta t_f$. In the $o$-th acoustic
substep, it is evaluated through linear temporal interpolation:
\begin{equation}
S_N^{\,t+o\Delta t}
=
S_N^{\,t}
+
\frac{o}{m}
\left(
S_N^{\,t+\Delta t_f}
-
S_N^{\,t}
\right),
\qquad
o=0,1,\ldots,m.
\label{eq:source_interpolation}
\end{equation}
The subcycling ratio is fixed at $m=1000$ in the present simulations.

\subsection{Solution Procedure}
The coupled hydrodynamic-acoustic solution is advanced using a
multirate, one-way coupling strategy. Within each hydrodynamic time
step, the cavitating-flow solution provides the base-state quantities
and acoustic source terms, while the CAPE system is advanced through
$m$ smaller acoustic substeps. The solution procedure is summarized as
follows:

\begin{enumerate}

\item Initialize the hydrodynamic variables
$(\boldsymbol{U},P,\alpha_l)$, the turbulence and cavitation-model
fields, and the acoustic variables
$(p',\rho',\boldsymbol{f})$. Specify the flow time step
$\Delta t_f$, the acoustic time step $\Delta t$, and the subcycling
ratio
\begin{equation*}
m=\frac{\Delta t_f}{\Delta t}.
\end{equation*}

\item Advance the incompressible cavitating-flow solution from $t$ to
$t+\Delta t_f$ using the PIMPLE pressure-velocity coupling algorithm.
The velocity, pressure, liquid volume fraction, turbulence variables,
and cavitation mass-transfer fields are updated within this step.

\item Evaluate the net cavitation mass-transfer rate
\begin{equation*}
\dot{m}=\dot{m}_c+\dot{m}_v,
\end{equation*}
and update the mixture properties $(\rho_0,\mu_{\mathrm{eff}})$ and
local sound speed. The hydrodynamic quantities required by the CAPE
solver,
\begin{equation*}
\left\{
P,\,
\frac{\partial P}{\partial t},\,
\rho_0,\,
\frac{\partial \rho_0}{\partial t},\,
\dot{m},\,
\boldsymbol{U}
\right\},
\end{equation*}
are stored at $t$ and $t+\Delta t_f$.

\item For each acoustic substep $o=1,2,\ldots,m$, evaluate the combined
source term $S_N^{\,t+o\Delta t}$ using the linear interpolation given
by Eq.~\eqref{eq:source_interpolation}. The required base-state
quantities are consistently interpolated at the same acoustic time
level.

\item Advance the CAPE solution from $t+(o-1)\Delta t$ to
$t+o\Delta t$ using the predictor-corrector procedure:

\begin{enumerate}
\item solve Eq.~\eqref{eq:co_velocity_predictor} for the predicted
co-velocity $\boldsymbol{f}^{*}$;

\item construct $\mathrm{\mathcal{H}}^{*}$ and solve
Eq.~\eqref{eq:acoustic_pressure_correction} for the corrected acoustic
pressure $p'^{\,*}$;

\item correct the co-velocity using
Eq.~\eqref{eq:co_velocity_correction} and reconstruct the corresponding
face fluxes;

\item repeat the pressure-co-velocity correction until the prescribed
residual tolerances are satisfied; two to four correction iterations
are generally sufficient;

\item update $\rho'$, $\boldsymbol{u}'$, and $c$ using
Eqs.~\eqref{eq:density_update},
\eqref{eq:acoustic_velocity_update}, etc., and apply the acoustic boundary
conditions and PML damping terms.
\end{enumerate}

\item After completion of all $m$ acoustic substeps, set
\begin{equation*}
t\leftarrow t+\Delta t_f,
\end{equation*}
and repeat Steps 2-5 until the prescribed final simulation time is
reached.
\end{enumerate}
This procedure preserves the one-way hybrid structure of the method: the cavitating-flow solution drives the acoustic field through the CAPE source terms, while the acoustic perturbations do not feed back into the hydrodynamic cavitation solution. It also preserves the time-scale separation between hydrodynamic evolution and acoustic propagation, which is essential for efficient hydroacoustic prediction in low-Mach-number cavitating flows.

\section{Results and Discussion}
The objective of this section is to verify the numerical properties of the proposed CAPE formulation and then to demonstrate its ability to resolve cavitation-induced acoustic source mechanisms in canonical multiphase flows. The analysis is carried out in three stages. First, one-dimensional acoustic propagation problems are used to isolate the numerical behavior of the solver, including mesh and time-step sensitivity, frequency preservation, absorbing-layer performance, and viscous acoustic attenuation. Second, flow past a circular cylinder is considered to establish the transition from non-cavitating dipole-dominated radiation to cavitation-driven monopole-like acoustic emission. Third, a cavitating hydrofoil is examined to assess whether the same collapse-induced acoustic source pathway persists in a lifting-body configuration, where the radiation pattern is additionally shaped by source localization and geometric shielding. Together, these cases provide both numerical verification and physical interpretation of the CAPE framework.

\subsection{Numerical Verification of One-Dimensional Wave Propagation}
\begin{figure}[h]
	\centering
	\centering\includegraphics[width=1.0\linewidth]{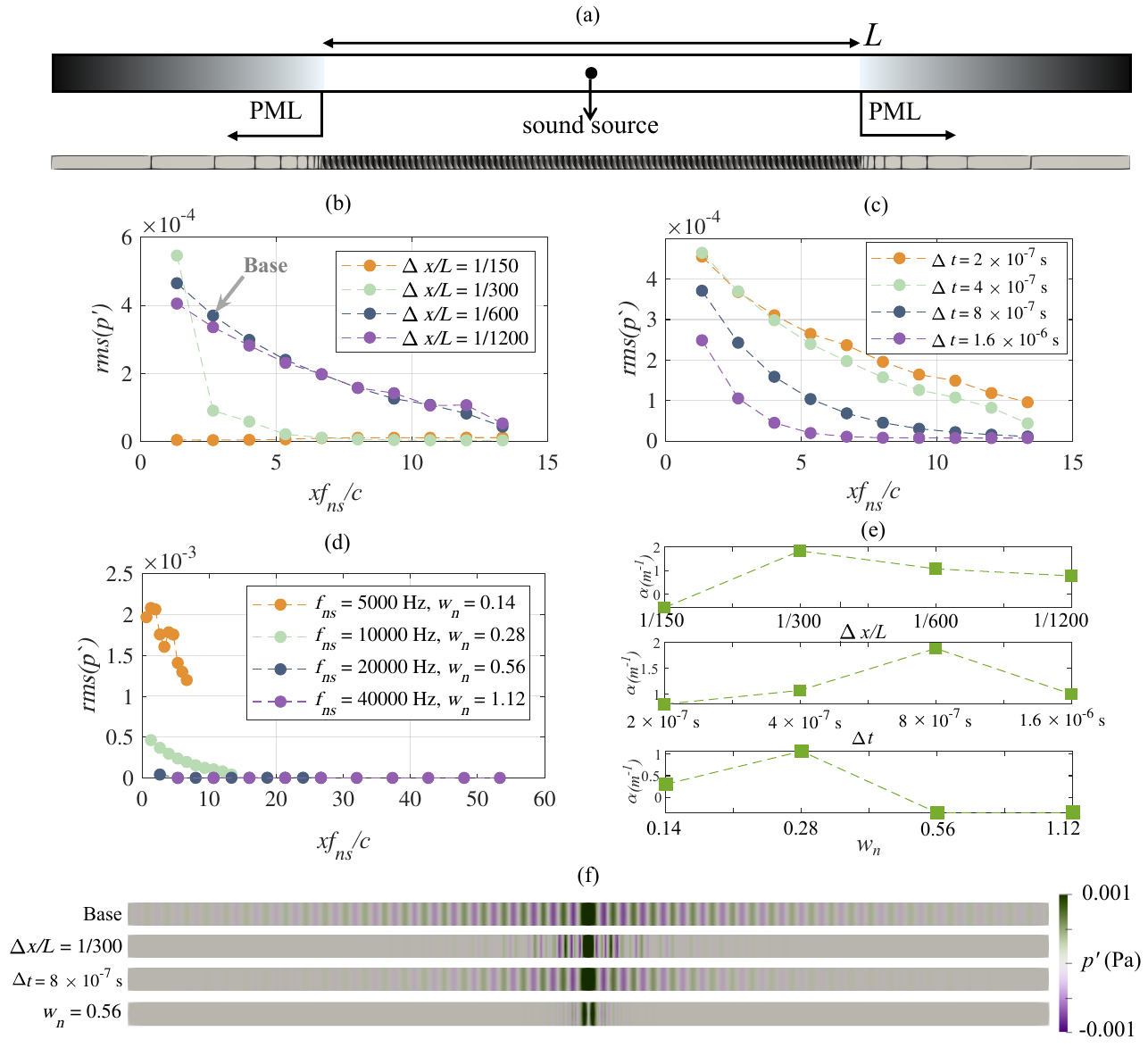}
	\caption{One-dimensional validation and sensitivity analysis of the present cavitation-aware acoustic solver with PML treatment. (a) Computational configuration with a harmonic point sound source located at the center of the domain and PML regions imposed near both boundaries. (b) Mesh sensitivity study of acoustic attenuation. (c) Time-step sensitivity study. (d) Effect of sound source frequency \(f_{ns}\), corresponding to different nondimensional wave numbers \(w_n\). (e) Attenuation coefficients extracted from subfigures (b-d). (f) Acoustic pressure contours of the selected reference case.}
	\label{rmsPp_alpha_mesh_deltaT_wn}
\end{figure}

To evaluate the numerical performance of the present cavitation-aware acoustic solver and the effectiveness of the PML treatment, a one-dimensional acoustic propagation problem is first considered, as illustrated in Fig.~\ref{rmsPp_alpha_mesh_deltaT_wn}(a). A harmonic point sound source is imposed at the center of the computational domain, while two PML regions with a length of \(L/2\) are applied near both boundaries to absorb outgoing acoustic waves. In all validation cases, the source terms \(h\), $\partial \rho_0 / \partial t$ and \(\partial P/\partial t\) in Eqs.~\ref{final_c_eq_0}-\ref{final_c_eq} are simplified as harmonic point sources with identical frequency \(f_{ns}\), while their amplitudes are prescribed as 20 (1/s) and 200 ($\rm m^2/s^3$), respectively. The acoustic attenuation coefficient within the PML region is set to \(\sigma_{max}=2\), and the mean flow velocity is set to zero within Fig.~\ref{rmsPp_alpha_mesh_deltaT_wn}.

Figures~\ref{rmsPp_alpha_mesh_deltaT_wn}(b-d) further investigate the influences of mesh resolution, physical time-step size, and frequency of sound source on acoustic attenuation behavior. In Fig.~\ref{rmsPp_alpha_mesh_deltaT_wn}(b), the mesh sensitivity study shows that coarse spatial discretization introduces noticeable numerical dissipation and inaccurate attenuation prediction, while the solution gradually converges as the mesh is refined from \(\Delta x/L=1/150\) to \(1/1200\). The base mesh exhibits sufficient resolution of the propagating acoustic wavelength. Figure~\ref{rmsPp_alpha_mesh_deltaT_wn}(c) presents the time-step sensitivity analysis. Large time steps generate excessive numerical damping and distort the attenuation process, whereas smaller time steps produce smoother and more stable wave propagation. Satisfactory temporal convergence is achieved when \(\Delta t \le 4\times10^{-7}\,\mathrm{s}\). Figure~\ref{rmsPp_alpha_mesh_deltaT_wn}(d) examines the influence of the frequency of the sound source \(f_{ns}\), which is directly correlated with the number of nondimensional waves \(w_n\) = $2\pi f_{ns} \Delta x/c$. As the frequency increases, the acoustic fluctuation attenuates more rapidly during propagation. This trend agrees with the physical mechanism of frequency-dependent sound attenuation, where high-frequency acoustic waves dissipate more rapidly than low-frequency components (investigated in detail later).

The attenuation coefficients extracted from Figs.~\ref{rmsPp_alpha_mesh_deltaT_wn}(b-d) are summarized in Fig.~\ref{rmsPp_alpha_mesh_deltaT_wn}(e). Fig.~\ref{rmsPp_alpha_mesh_deltaT_wn}(f) presents the acoustic pressure contours of the selected reference case with the corresponding parameters changed. Smooth wave propagation and gradual attenuation are observed throughout the computational domain without noticeable spurious reflections near the boundaries.

\begin{figure}[h]
	\centering
	\centering\includegraphics[width=1.0\linewidth]{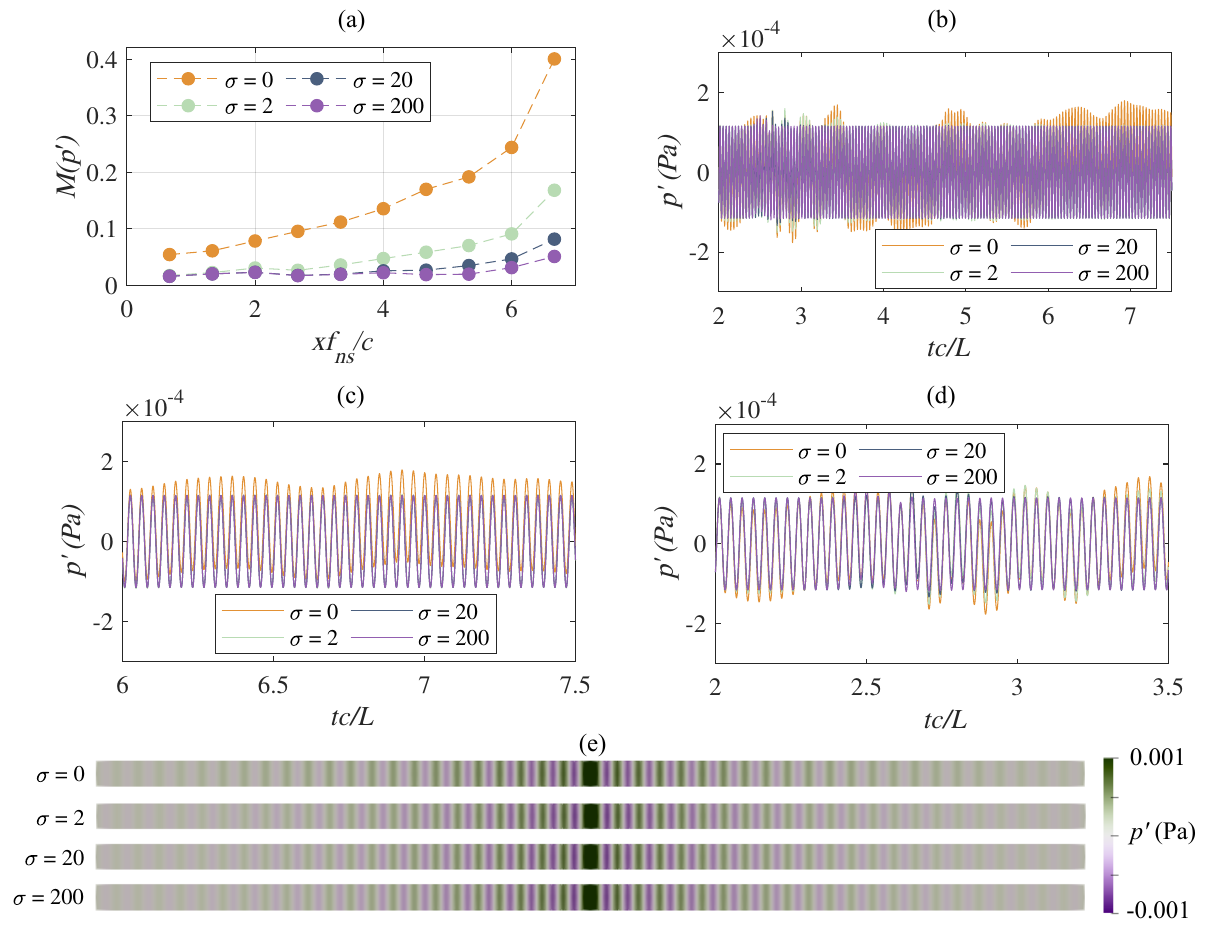}
	\caption{Sensitivity analysis of the PML damping amplitude \(\sigma\) for the one-dimensional acoustic propagation problem. (a) Modulation index \(M(p')\) along the propagation direction, quantifying reflection-induced envelope modulation. (b) Acoustic pressure fluctuations measured at $xf_{ns}/c=6$. (c,d) Enlarged local temporal windows of panel (b). (e) Acoustic pressure contours for different damping amplitudes, showing the suppression of standing-wave contamination after applying the PML treatment.}
	\label{MPp_vs_varyingpml}
\end{figure}

Figure \ref{MPp_vs_varyingpml} investigates the influence of the PML damping amplitude \(\sigma\) on the non-reflective performance of the present acoustic solver. In the present test, the mean flow velocity is fixed at \(0~\mathrm{m/s}\), while \(\Delta x/L=1/600\), \(\Delta t=4\times10^{-7}~\mathrm{s}\), and \(w_n=0.28\) remain unchanged. Four damping amplitudes are considered, including \(\sigma_{max}=0\) (without PML), 2, 20, and 200. Figure \ref{MPp_vs_varyingpml}(a) presents the modulation index \(M(p')\), which quantifies the reflection-induced modulation of the amplitude of the acoustic signal:
\[
M(p')=\frac{sd(A(t))}{\overline{A(t)}},
\]
where \(A(t)\) is the envelope of the acoustic pressure fluctuation $p'$ extracted using the Hilbert transform, and \(sd(\cdot)\) represents the standard deviation operator and \(\overline{A(t)}\) denotes the temporal average of the envelope amplitude. Without PML, the modulation level continuously increases during propagation due to strong wave reflections and standing-wave contamination. After introducing PML treatment, modulation is significantly suppressed, while further increasing \(\sigma_{max}\) beyond 2 only produces marginal improvements. The pressure fluctuations measured at $xf_{ns}/c=6$ in Fig.~\ref{MPp_vs_varyingpml}(b), together with the enlarged local temporal windows in Figs.~\ref{MPp_vs_varyingpml}(c,d), further demonstrate that the reflected-wave-induced envelope modulation disappears after applying the PML treatment, resulting in nearly monochromatic harmonic oscillations. Finally, the acoustic pressure contours in Fig.~\ref{MPp_vs_varyingpml}(e) show that obvious standing-wave structures exist without PML, whereas smooth outgoing wave propagation and gradual attenuation are achieved once the PML region is introduced. Overall, the present results indicate that a moderate damping amplitude of \(\sigma_{max}=2\) is already sufficient to provide satisfactory non-reflective performance for the current simulations.

\begin{figure}[h]
	\centering
	\centering\includegraphics[width=1.0\linewidth]{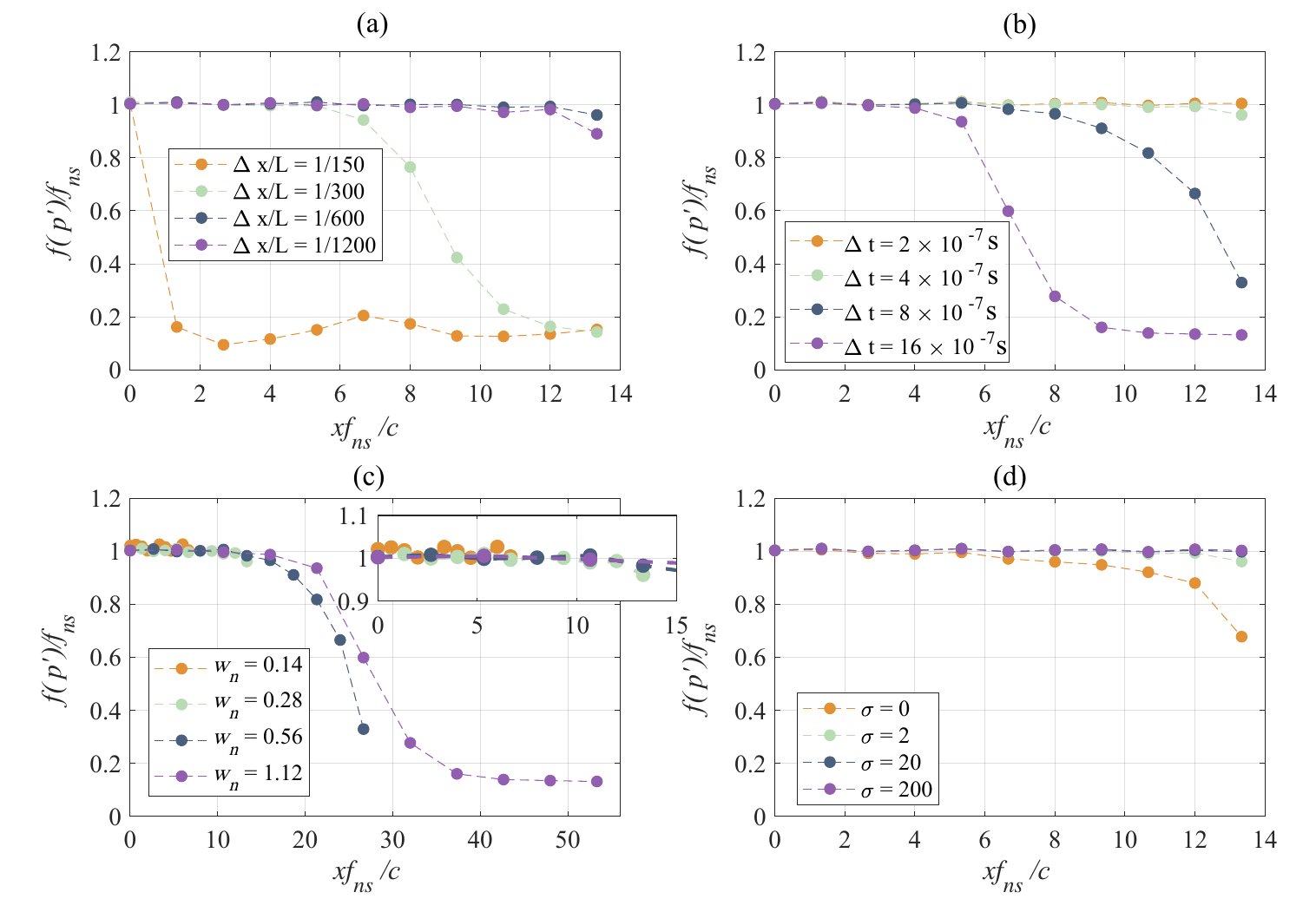}
	\caption{Frequency preservation study of the one-dimensional acoustic propagation problem. 
Normalized dominant acoustic frequency, $f(p')/f_{ns}$, along the propagation direction under varying (a) mesh resolutions, (b) time-step sizes, (c) nondimensional wave numbers $w_n$, and (d) PML damping amplitudes $\sigma$.}
	\label{FrePp_indenpency}
\end{figure}
Figure~\ref{FrePp_indenpency} presents the frequency dependency study for different numerical parameters. In contrast to the attenuation of the amplitude observed in Figs.~\ref{rmsPp_alpha_mesh_deltaT_wn} and \ref{MPp_vs_varyingpml}, the dominant pressure frequency is generally preserved over a much longer propagation distance, with $f(p')/f_{ns}$ remaining close to unity for sufficiently refined discretizations. As shown in Figs.~\ref{FrePp_indenpency}(a,b), the resolution of the mesh and the size of the time steps have the strongest influence on frequency preservation, and the frequency deviations appear only when relatively coarse meshes or large time steps are employed. The wave number $w_n$ affects the frequency prediction only for excessively large values, while the influence of the damping amplitude $\sigma$ remains negligible throughout most of the computational domain. These results indicate that the proposed method is capable of maintaining the dominant acoustic frequency with good accuracy, provided that appropriate spatial and temporal resolutions are used.

\begin{figure}
	\centering
	\centering\includegraphics[width=1.0\linewidth]{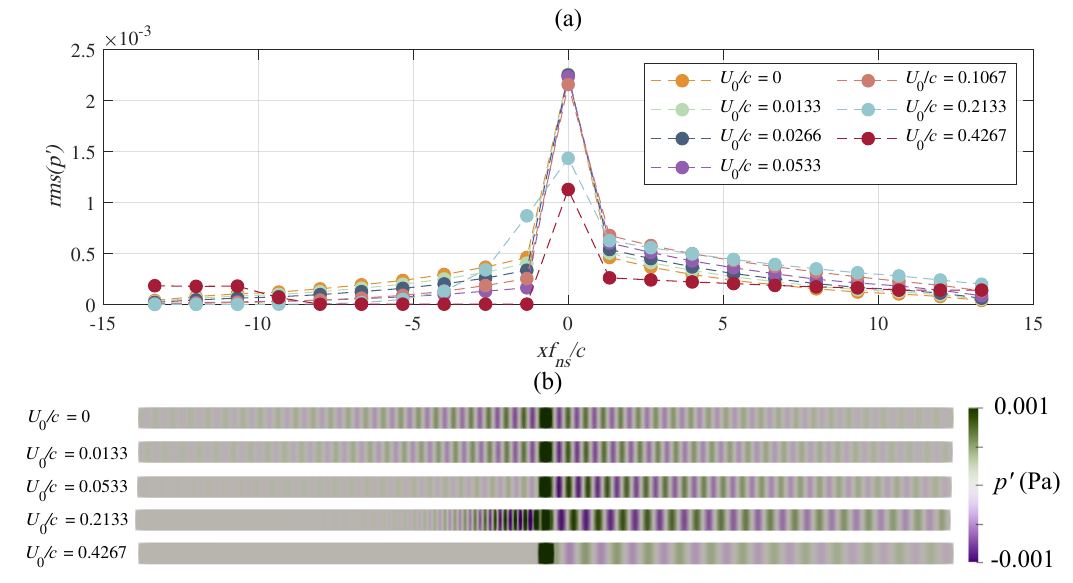}
	\caption{Influence of mean-flow velocity on acoustic propagation. (a) Spatial distribution of \(rms(p')\) under different nondimensional inflow velocities \(U_0/c\). (b) Corresponding acoustic pressure contours, showing convection-induced asymmetric wave propagation.}
	\label{rmsPp_vs_velocity}
\end{figure}

Figure \ref{rmsPp_vs_velocity} further examines the influence of mean-flow convection on the behavior of acoustic propagation. In the present test, spatial resolution, physical time step, nondimensional wave number and PML damping amplitude are fixed at \(\Delta x/L=1/600\), \(\Delta t=4\times10^{-7}~\mathrm{s}\), \(w_n=0.28\), and \(\sigma_{max}=2\), respectively, while nondimensional inflow velocity \(U_0/c\) varies from 0 to 0.4267. Figure \ref{rmsPp_vs_velocity}(a) presents the spatial distribution of the intensity of the fluctuation of acoustic pressure, quantified using \(rms(p')\). As the mean-flow velocity increases, the downstream acoustic fluctuation is first enhanced due to the convective transport of acoustic energy by the background flow. However, when the inflow velocity becomes sufficiently large, the downstream fluctuation amplitude gradually decreases because the acoustic wave rapidly convected away from the source region, reducing the local accumulation of acoustic energy. Meanwhile, the upstream acoustic fluctuation is continuously suppressed with increasing inflow velocity, indicating the weakening of upstream wave propagation under strong convective effects. Figure \ref{rmsPp_vs_velocity}(b) further presents the corresponding acoustic pressure contours. The acoustic wavelength becomes increasingly stretched in the downstream direction and compressed upstream as the mean flow increases, demonstrating the convection-induced asymmetry of acoustic propagation under moving background flow conditions.
These tests establish the numerical resolution and PML parameters used in the following. The physical attenuation predicted by the viscous terms is
examined separately against Stokes' law in the next subsection.

\subsection{Verification against Stokes' Attenuation Law}

\begin{figure}[ht!]
	\centering
	\centering\includegraphics[width=1.0\linewidth]{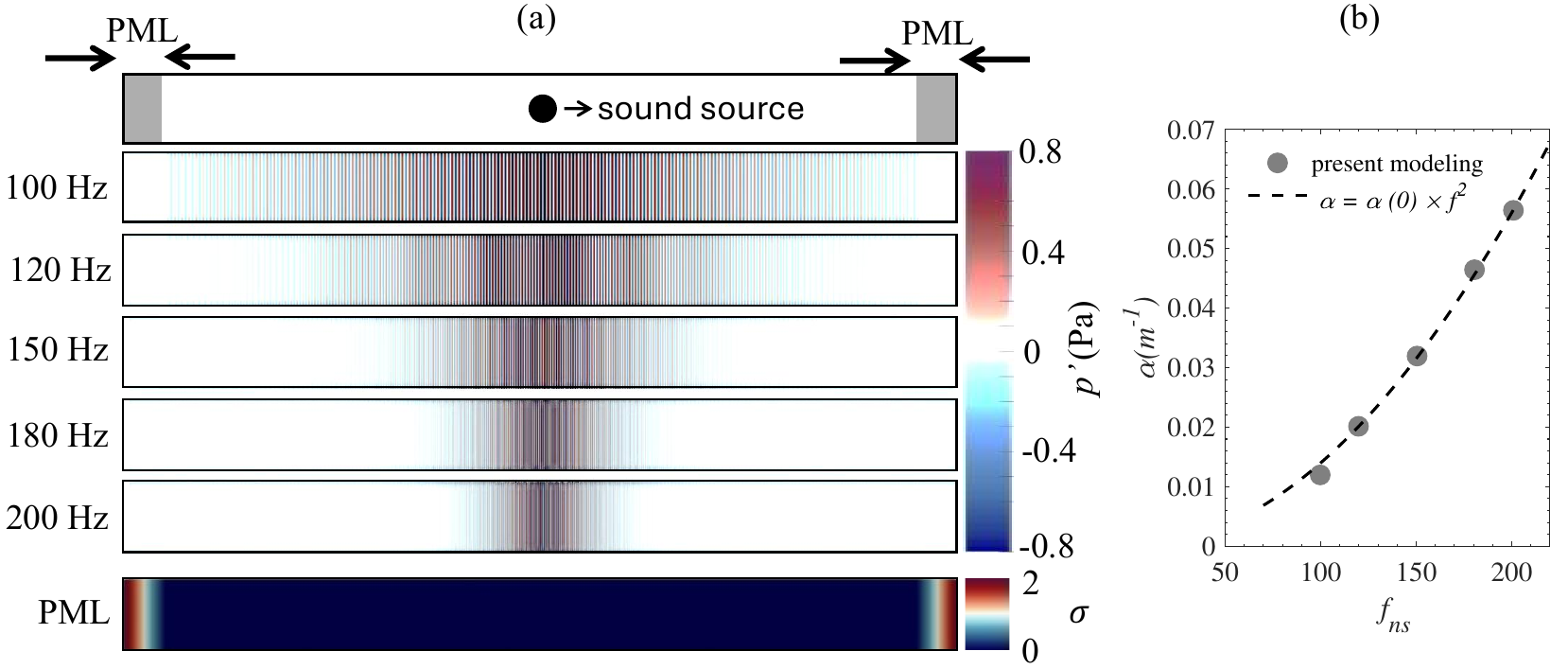}
	\caption{Stokes’s law validation of acoustic propagation and attenuation. (a) Contours of fluctuating pressure generated by a harmonic point source for frequencies from 100~Hz to 200~Hz, with perfectly matched layers (PML) applied at both ends of the domain. (b) Extracted attenuation coefficient as a function of frequency.}
	\label{stocks_law}
\end{figure}

Stokes' law of sound attenuation is a formula for acoustic attenuation in Newtonian fluid, such as water or air, due to the viscosity of the fluid \cite{stokes_2009}. In an isotropic and homogeneous Newtonian medium, Stokes' law of sound attenuation is shown as:
\begin{linenomath}
\begin{equation}\label{eq:ns}
p'\left( d \right) =p_0'e^{-\alpha d}.
\end{equation}
\end{linenomath}

After traveling a distance \(d\), the fluctuating pressure amplitude \(p_0'\) will attenuate to \(p'\left( d \right) \). The attenuation coefficient \(\alpha \,\,\left( m^{-1} \right) \) is proportional to the square of the frequency of the noise source, as \(\alpha \propto f_{ns}^2\).
Here, the propagation of the sound source is studied in a one-dimensional problem to verify the present APE framework. As shown in Fig. \ref{stocks_law}a, a point sound source caused by cavitation-induced regular collapse is set in the geometric center of the domain with a normalized side length (= $ L/c \Delta t $) of 50,000 in direction $x$, $L$ is the physical side length. 
Here, the sound propagation of frequencies ranging from 100 Hz to 200 Hz is analyzed. 
Since Stokes’ Attenuation Law is valid in a homogeneous medium, we here treat all variables related to cavitation as vacant in this validation case.
Equation \ref{final_c_eq} is changed into:
\begin{linenomath}
\begin{equation}\label{eq:ns3}
\\
\frac{\partial p'}{\partial t}+c \sigma  p'-c^2\frac{\partial \rho '}{\partial t}=-\frac{\partial \left( p_0'\cos \left( 2\pi ft \right) \right)}{\partial t}=2\pi f_{ns}\cdot p_0'\sin \left( 2\pi f_{ns}t \right),
\\
\end{equation}
\end{linenomath}
with the amplitude of the source pressure \(p_0'\) at the source location. The velocity of the computational domain is set to zero and the physical time step $\Delta t$ is 1$\times 10^{-5}$ s.

Fig. \ref{stocks_law}a shows the contour of the fluctuating pressure propagating outside with varied frequencies. As can be seen, the higher the source frequency, the closer the pressure extremum appears in the source location, and the faster the noise is attenuated in the propagation medium.
Fig. \ref{stocks_law}b shows the attenuation exponents (referred to the general attenuation from 5000 to 12,500) as a function of frequencies. The present results of noise attenuation satisfy \(\alpha \propto f_{ns}^2\), which is consistent with Stokes' law of sound attenuation. This agreement verifies that the viscous terms in the APE solver reproduce the correct frequency-dependent attenuation trend for a Newtonian medium.

\subsection{Non-Cavitating Cylinder Benchmark}

\begin{figure}[h]
	\centering
	\centering\includegraphics[width=1.0\linewidth]{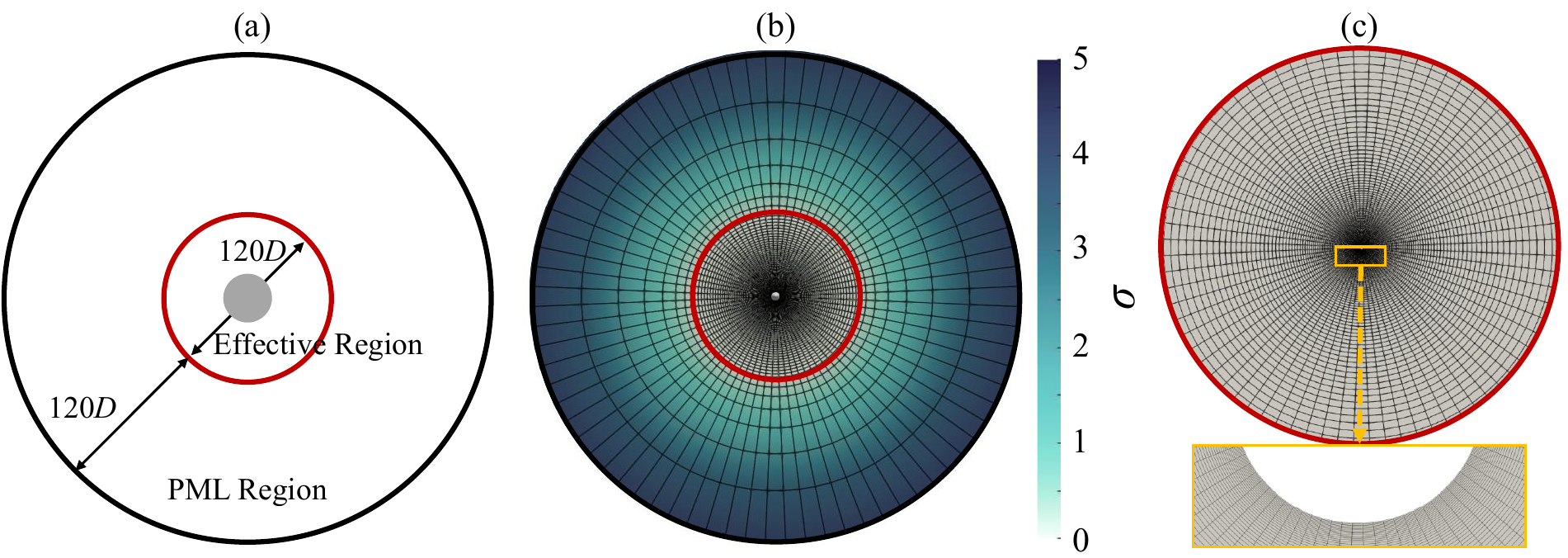}
	\caption{Computational domain and mesh configuration: (a) Schematic of the circular domain, showing the effective region surrounding the body and the outer perfectly matched layer (PML) used to absorb outgoing acoustic waves; the radii of both regions are indicated in terms of the characteristic length diameter $D$. (b) Full computational mesh with the color map indicating the spatial distribution of the damping coefficient $\sigma$ within the PML region. (c) Enlarged view of the near-field mesh around the body, highlighting the local grid refinement used to accurately resolve cavitation-induced acoustic sources; the inset shows the detailed mesh near the solid surface.}
	\label{mesh_and_PML_domain}
\end{figure}

We also consider examples of flow passing past a circular cylinder, and a single-phase configuration at $Re$ = 200 is explored first.
Figure~\ref{mesh_and_PML_domain} illustrates the computational domain used to simulate sound generation and propagation induced by unsteady flow past a circular cylinder. 
To accurately capture both near-field flow-acoustic sources and far-field noise radiation, the domain is divided into an inner Effective Region and an outer Perfectly Matched Layer (PML). As shown schematically in Fig.~\ref{mesh_and_PML_domain}(a), the effective region, which surrounds the cylinder, the surrounding flow field, and the acoustic sources, extends radially to $120D$, where $D$ is the diameter of the cylinder. Beyond this region, an annular PML with a thickness of $120D$ is introduced to absorb outgoing acoustic waves and prevent artificial reflections at the outer boundary.

Figure~\ref{mesh_and_PML_domain}(b) presents the spatial distribution of the PML damping coefficient $\sigma$, which increases smoothly from zero at the interface between the Effective Region and the PML to its maximum value at the far-field boundary. This gradual
variation ensures impedance matching between the physical domain and the absorbing layer, enabling efficient attenuation of outgoing acoustic waves over a wide range of frequencies without generating spurious reflections. 
The computational domain is covered by an O-mesh, with mesh quality coarsened in the PML region to dissipate acoustic energy. Figure~\ref{mesh_and_PML_domain}(c) shows the computational mesh used in the Effective Region. The grid is refined near the cylinder and is progressively coarsened toward the outer boundary, while sufficient resolution is maintained to ensure the resolution of the boundary layer.

As illustrated in Fig. \ref{mesh_and_PML_domain}(c), the computational domain is discretized using an O-type mesh with uniform spacing in the circumferential direction and progressive
stretching in the radial direction.
The corresponding lift, drag coefficients and the Strouhal number of vortex-shedding are calculated at different mesh resolutions, and the associated results are summarized in Table \ref{mesh_com}.
It could be observed that the relative differences of each parameter between mesh 1 to mesh 2 are considerable, but all decrease to a smaller value as the mesh is refined to mesh 3 (fine) and mesh 4 (very fine).
As a consequence, the strategy applied by mesh 3 is adopted in all configurations of the present work to achieve the best balance of calculation time and accuracy.

\begin{figure}
	\centering
	\centering\includegraphics[width=0.9\linewidth]{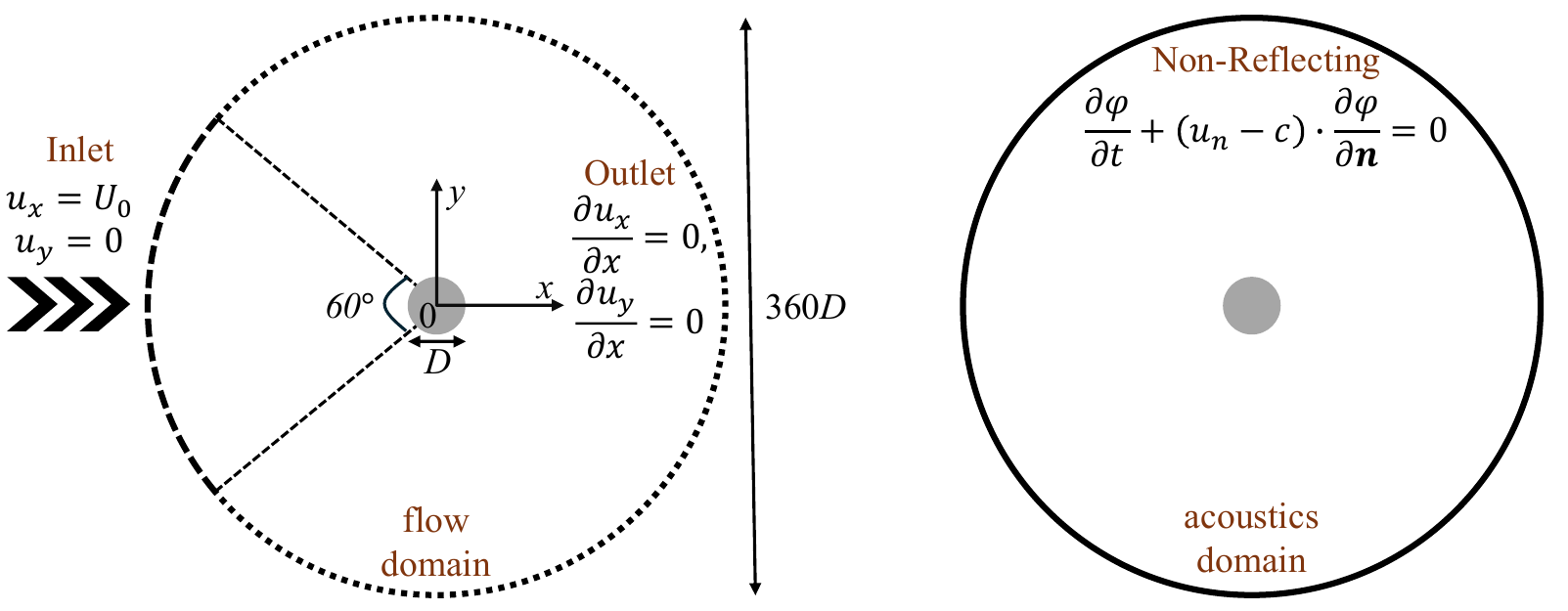}
	\caption{Schematic of the coupled flow and acoustic computational domains. The flow domain (left) shows the inlet velocity condition, outlet zero-gradient conditions, and geometric configuration relative to the characteristic length $D$. The acoustic domain (right) is used to solve the acoustic perturbation equations with a non-reflecting boundary condition imposed at the outer boundary, allowing outgoing acoustic waves to exit the domain without spurious reflections.}
	\label{flow_noise_domain}
\end{figure}

Figure~\ref{flow_noise_domain} further details the boundary conditions used for hybrid flow and acoustic simulations. For the flow domain, shown on the left, a uniform inflow condition is prescribed at the input, with $u_x = U_0$ and $u_y = 0$, while the outflow boundary applies a zero-gradient condition, $\partial u_x / \partial x = 0$ and $\partial u_y / \partial x = 0$, allowing vortical structures to convect smoothly out of the domain.  For the acoustic domain, at the outer boundary, the above-introduced non-reflecting boundary condition is imposed on the outer edge of the PML region. This combination allows incoming acoustic waves to exit the domain without spurious reflection, effectively emulating an unbounded acoustic medium.

\begin{table*}
\centering
\caption{\textcolor{black}{Mesh convergence study: Lift coefficients $C_{L,rms}$, drag coefficients $C_{D,rms}$ and reduced vortex shedding frequency $F^*_{vs}$. $\Delta \hat{x}_{min} $ is the minimum relative grid length.
Effective region represents the red-marked region in Fig. \ref{mesh_and_PML_domain}.}
}
\begin{tabular}{  c  c  c  c  c  c  c  c  c  c  c }
\hline
Mesh &  Cells Number & Effective Region & $C_{L,rms}$ & $C_{D,rms}$  & $F^*_{vs}$ & $\Delta \hat{x}_{min}$ & $y^+$ \\
\hline

 1 & 7,424 & 6,976 & 1.2934 & 0.1694 & 0.367 & 0.0070 & 5 \\ 
 2 & 9,520 & 9,030 & 1.4986 & 0.1812 & 0.348 & 0.0040 & 3 \\ 
 3 & 11,856 & 11,324 & 1.5823 & 0.1944 & 0.335 & 0.0025 & 2 \\ 
 4 & 14,432 & 13,858 & 1.5910 & 0.1989 & 0.339 & 0.0015 & 1 \\ 
\hline
\end{tabular}
\label{mesh_com}
\end{table*}

\begin{figure}[h]
	\centering
	\centering\includegraphics[width=0.7\linewidth]{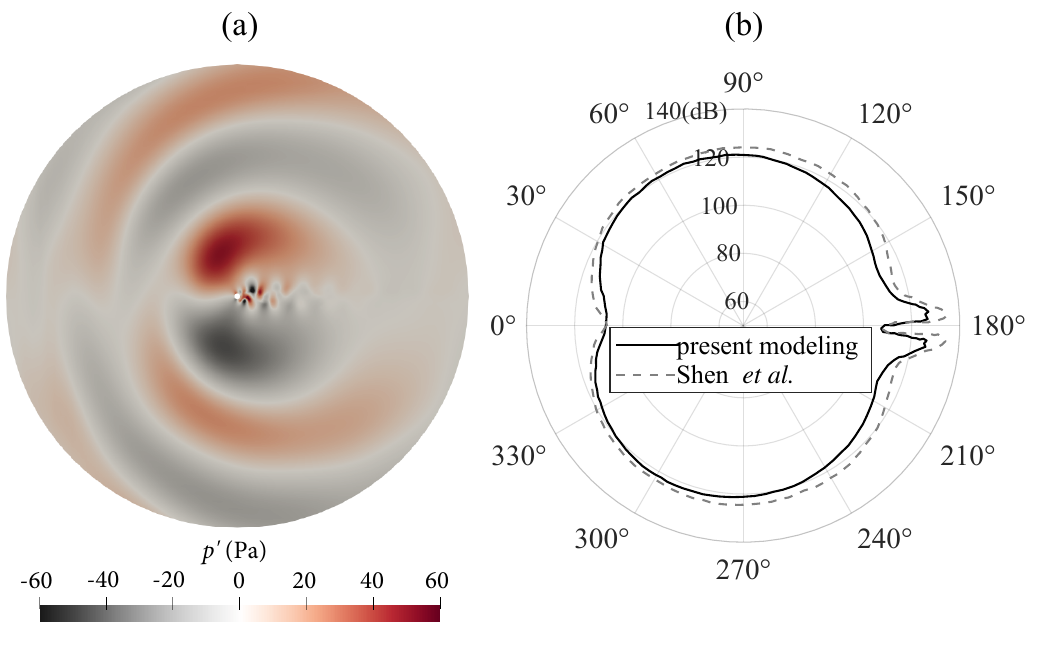}
	\caption{Validation of the hybrid acoustics model for flow-induced noise from a circular cylinder. (a) Instantaneous fluctuating acoustic pressure field at $Re=200$ after reaching a statistically stationary state. (b) Sound pressure level ($SPL$) directivity at a radial distance of $r=20D$ at $Re=200$, compared with results from \citet{shen2004collocated}. }
	\label{cylinder_re200_COM}
\end{figure}

For the present experimental case at $Re$ = 200, Fig. \ref{cylinder_re200_COM}a shows the fluctuating pressure of the sound wave after the acoustic field has become statistically stationary. The sources of the sound shown here include the pressure fluctuations on the cylinder and the wake vortex street. The fluctuating pressure associated with the sound wave propagates to the far field of the computational domain.
Then the distribution \(SPL\) is analyzed, with \(P_{ref}\) as the reference pressure of \(2\times 10^{-5} Pa\) (an environment of aerodynamics is assumed here for comparison with the reported data) and \(P_e\) as the root mean square value of fluctuating pressure. Here, the time range \(t_1\) to \(t_2\) including about 50 fluctuating periods is chosen for analysis:

\begin{linenomath}
\begin{equation}
P_e=\sqrt{\frac{1}{t_2-t_1}\int{\begin{array}{c}
	t_2\\
	t_1\\
\end{array}\left[ p'\left( t \right) \right] ^2dt}},
\end{equation}
\end{linenomath}

\begin{linenomath}
\begin{equation}
SPL=20\log _{10}\frac{P_e}{P_{ref}},
\end{equation}
\end{linenomath}

Fig. \ref{cylinder_re200_COM}b shows the $SPL$ distribution as a function of the azimuth angle $q$ at a radial distance of $r = 20D$ from the center of the cylinder (directivity pattern). The results of the present simulations are compared with similar results obtained by \citet{shen2004collocated}. In general, the present predictions of directivity $SPL$ agree very well with those reported by \citet{shen2004collocated}. The directivity pattern for $SPL$ exhibits high pressure levels transverse to the incident flow direction, corresponding to the vortex sheet that the circular cylinder sheds into the wake on the alternating sides of the cylinder. The $SPL$ achieves a minimum at the stagnation point and at the rear point of the cylinder, respectively. Note that the $SPL$ directivity pattern exhibits symmetry on the horizontal axis.

\subsection{Cavitating Flow Passing Cylinder at High-Reynolds Number Flow}
\label{}

In order to investigate more closely the effect of cavitation on noise generation using the present framework, in this section, the flow is modeled as a two-phase system consisting of water and vapor, both treated
as Newtonian fluids. 
The water phase is assigned a density of
$\rho = 1000~\mathrm{kg\,m^{-3}}$ and a kinematic viscosity of
$\nu = 9\times10^{-7}~\mathrm{m^2\,s^{-1}}$, while the vapor phase has a density of
$\rho = 0.02308~\mathrm{kg\,m^{-3}}$ and a kinematic viscosity of
$\nu = 4.273\times10^{-4}~\mathrm{m^2\,s^{-1}}$.
The mean water inflow velocity $U_0$ is 5 $\rm ms^{-1}$ with Reynolds number $Re = 1.11 \times 10^5$. The circular cylinder has a diameter of 0.02 $\rm m$. The level of free-stream turbulence was 2\%. The free-stream turbulence is an important quantity for acoustics because increased free-stream turbulence can increase the noise level.
The cavitation number is defined as $
\sigma_{\mathrm{cav}}
= (p_{\mathrm{ref}} - p_{\mathrm{vap}})/(\tfrac{1}{2}\rho U_\infty^2) $, where $p_{\mathrm{ref}}$ is the reference pressure and $p_{\mathrm{vap}}$ is the saturation pressure. $\sigma_{\mathrm{cav}}$ in the present work is set as 0.8 .
The domain, mesh, and boundary conditions are consistent with those described in Figs .\ref{flow_noise_domain} and \ref{mesh_and_PML_domain}.

\begin{figure}
	\centering
	\centering\includegraphics[width=1.0\linewidth]{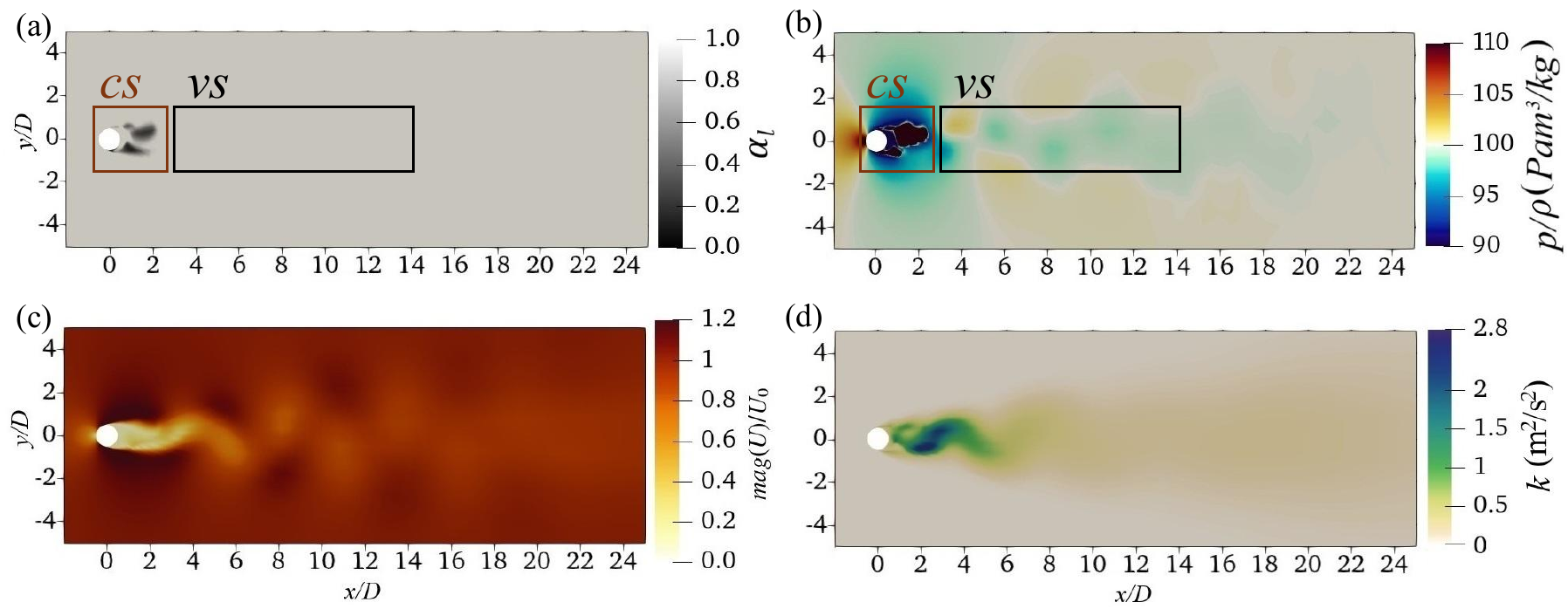}
	\caption{Flow-field characteristics for cavitation flow past a circular cylinder. (a) Phase indicator field $\alpha_l$. (b) Instantaneous kinematic pressure, $p/\rho$. (c) Normalized velocity magnitude, $mag{(U)}/U_0$. (d) Turbulent kinetic energy, $k$.}
	\label{cylinder_flow_fields}
\end{figure}

When the vortex-shedding with cavitation generation reaches a periodic equilibrium state, we exhibit in Fig. \ref{cylinder_flow_fields} the instantaneous contours of cavitation (with volume fraction of 0.9), kinematic pressure ($p/\rho$), normalized velocity magnitude $mag(U)/U_0$, and Turbulent kinetic energy $k$.
In scenarios in which the flow passes a circular cylinder without cavitation, vortex-shedding behaviors will be the most significant noise sources. In the present work, as we consider the impact of cavitation, the wake area in Fig. \ref{cylinder_flow_fields}a is therefore divided into two regions, dominated by cavitation-shedding ($cs$) and vortex-shedding ($vs$), respectively.
In Fig. \ref{cylinder_flow_fields}b, the pressure in the cavitation region is extremely low due to the presence of vapor. This further impacts the wake, resulting in a pressure distribution in the $vs$ region that is significantly different from the non-cavitating case, exhibiting a more spatially divergent pattern. This is also further demonstrated by the velocity and kinetic-energy contour in figures \ref{cylinder_flow_fields}c and d.

\begin{figure}
	\centering
	\centering\includegraphics[width=1.0\linewidth]{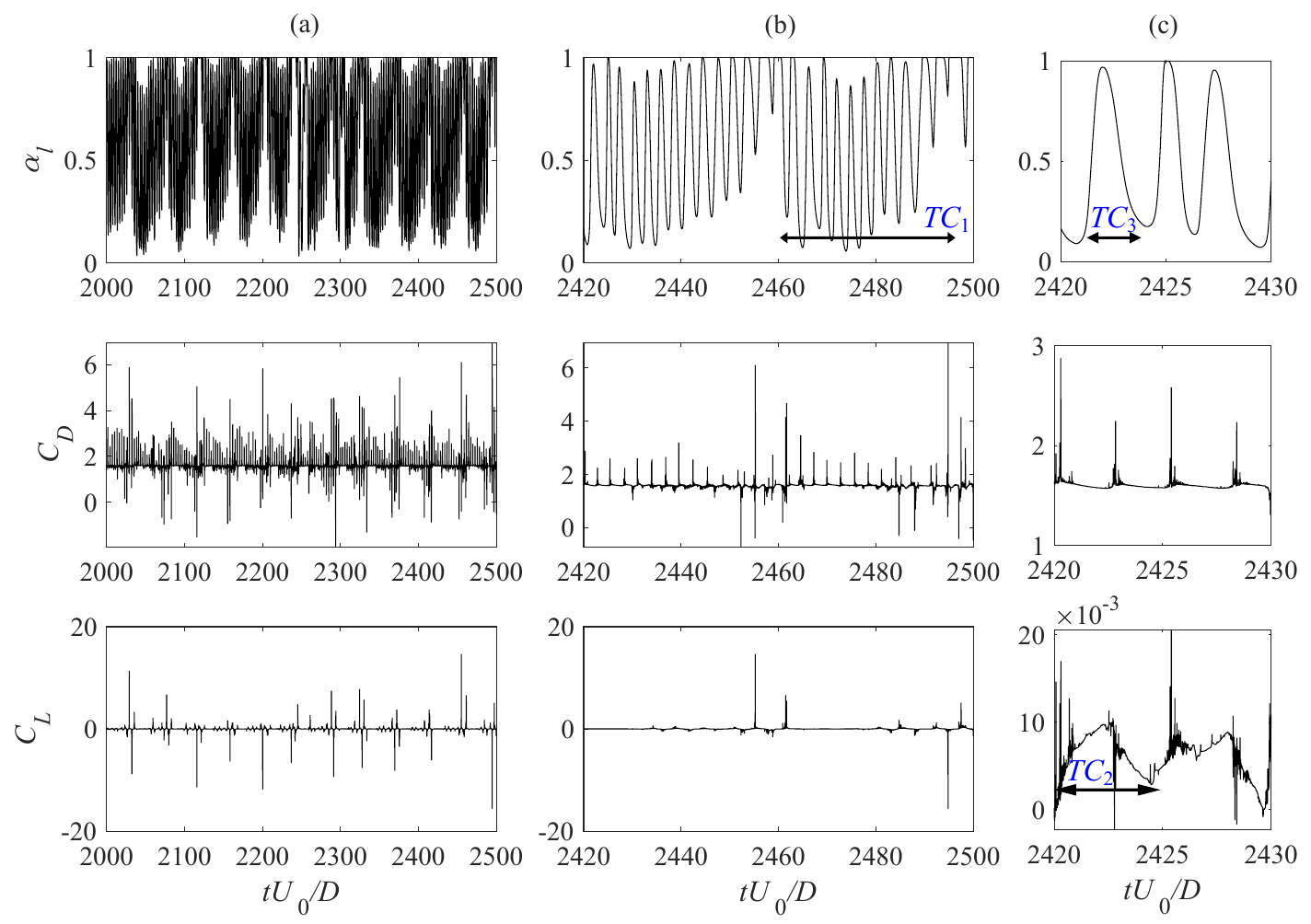}
	\caption{Instantaneous time histories of local liquid volume fraction $\alpha_l$, drag coefficient $C_D$, and lift coefficient $C_L$ over three progressively refined temporal windows from panels (a) to (c), illustrating multiscale cavitation and force dynamics.}
	\label{alpha_Cd_CL_1}
\end{figure}

Figure~\ref{alpha_Cd_CL_1} presents the instantaneous time histories of the local liquid volume fraction $\alpha_l$ at ($D$, 0), the drag coefficient $C_D$, and the lift coefficient $C_L$ on three progressively refined temporal scales. During the long window (panel~(a)), all signals exhibit strongly intermittent and impulsive behavior, reflecting the unsteady nature of cloud cavitation in the wake of the cylinder. At an intermediate scale (panel~(b)), a slower modulation with period $TC_1$ becomes evident, appearing as an envelope that regulates the intensity of the faster oscillations in $\alpha_l$ and the associated force fluctuations. This low-frequency component is attributed to a global cavity `breathing' mode \citep{Lak_Jaiman_2024}, corresponding to the gradual growth and collapse of the cavitating wake, which intermittently amplifies or suppresses the shedding dynamics. On the shortest scale (panel~(c)), two smaller time scales are clearly identified: $TC_2$, associated with the vortex-shedding process that governs the oscillatory lift response, and $TC_3$, corresponding to local cavity-shedding or collapse events observed in the vapor fraction signal. Together, these results highlight the coexistence and coupling of multiple characteristic time scales in cavitating flow past a circular cylinder, with the low-frequency and high-frequency modulations exerting a controlling influence on both vortex and cavity-shedding processes.

In addition, shedding and subsequent collapse of the cavity generate strong, periodic dynamic impulses on the surface of the cylinder, which manifest themselves as impulsive contributions to both lift and drag forces. These force peaks are temporally correlated with rapid variations in the local liquid volume fraction, indicating that cavitation-induced pressure transients play a dominant role in instantaneous loading. In particular, the drag response is largely governed by cavity dynamics, as evidenced by the close correspondence between high-$\alpha_l$ events and pronounced peaks in $C_D$, while lift fluctuations remain influenced by both cavity shedding and the underlying vortex-shedding asymmetry in the wake. This distinction highlights the primary role of cavitation in setting the magnitude of the streamwise force, while transverse loading reflects the coupled interaction between cavitation and wake instability.

\begin{figure}[h]
	\centering
	\centering\includegraphics[width=0.65\linewidth]{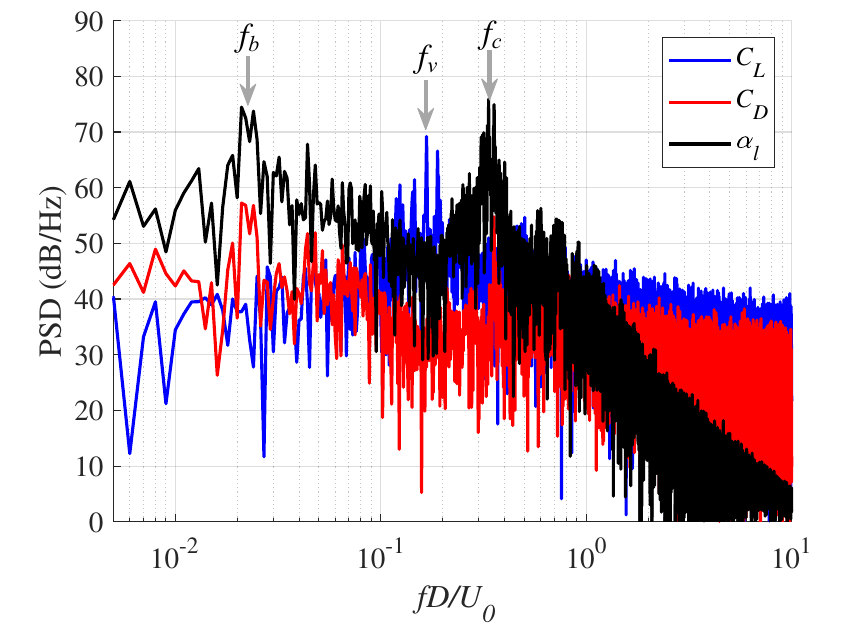}
	\caption{Power spectral density of $\alpha_l$, $C_D$, and $C_L$, showing the dominant characteristic frequencies associated with cavity breathing, vortex shedding, and cavity shedding.}

	\label{cl_cd_slpha_PSD}
\end{figure}

The three dominant spectral peaks identified in Fig.~\ref{cl_cd_slpha_PSD} are directly linked to the characteristic time scales observed in the time-domain signals. Specifically, the low-frequency peak $f_b$ of 0.023 corresponds to the long-period modulation $TC_1 \,(=1/f_b)$ identified in panel~(b) of Fig.~\ref{alpha_Cd_CL_1}, and is associated with the global `breathing' mode of the cavitating wake, characterized by slow growth and collapse of the cavity. The intermediate peak $f_v$ of 0.195 corresponds to the time scale $TC_2 \,(=1/f_v)$ and is most pronounced in the lift spectrum, consistent with the `vortex-shedding' process governing wake asymmetry and transverse force oscillations.  This indicates that transverse force fluctuations remain governed by the classical vortex-shedding frequency (with a Strouhal number near 0.2) even under cavitating conditions. The higher-frequency peak $f_c$ of 0.335 corresponds to the shortest time scale $TC_3 \,(=1/f_c)$ and is dominant in the vapor fraction spectrum, reflecting periodic shedding and collapse of the cavity. These cloud-collapse events generate impulsive pressure loads on the cylinder surface, which contribute strongly to drag fluctuations and to the broadband enrichment of the force spectra. Together, the correspondence between $(f_b,f_v,f_c)$ and $(TC_1,TC_2,TC_3)$ confirms the coexistence of multiple hierarchically coupled time scales in the cavitation flow past a circular cylinder. These high-frequency cavitation-induced impulses strongly contribute to the drag fluctuations and broaden the force spectra over a wide frequency range.

\begin{figure}[h]
	\centering
	\centering\includegraphics[width=1.0\linewidth]{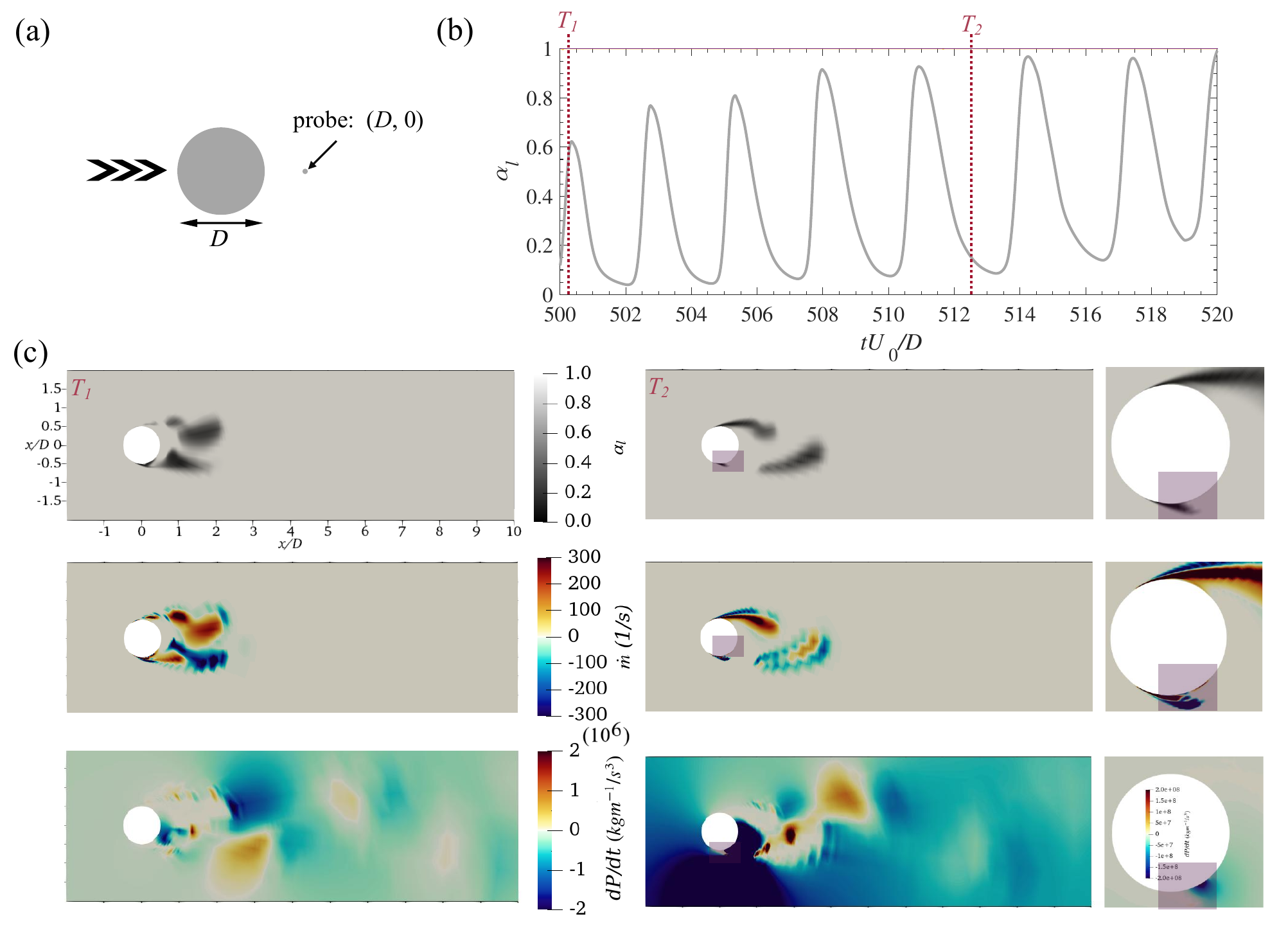}
	\caption{Conditional visualization of cavitation dynamics at two representative instants $T_1$ and $T_2$: (a) probe location, (b) local vapor fraction history, and (c) instantaneous contours of vapor fraction, cavitation mass-transfer rate, and source pressure time derivative.}
	\label{alpha_time_point_2.001_2.05s}
\end{figure}

In this case, to further elucidate the localized flow structures associated with cloud release and collapse, Fig.~\ref{alpha_time_point_2.001_2.05s} presents a conditional visualization of cavitation dynamics at two representative instants marked by $T_1$ and $T_2$. Panel~a illustrates the location of the probe at $(D,0)$ near the cylinder wake, where the vapor fraction signal is sampled. The corresponding time history of the local liquid volume fraction $\alpha_l$ is shown in panel~(b), with the instants $T_1$ and $T_2$ indicated to represent distinct phases within a cavitation cycle. Panel~c compares the instantaneous contours of the vapor fraction $\alpha_l$, the cavitation mass-transfer rate $\dot{m}$ obtained from the cavitation transport equation \ref{eq:masstransfer}, and the local pressure time derivative $dP/dt$ at these two instants. At $T_1$, the flow is characterized by the presence of a recently detached vapor cloud accompanied by intense localized condensation and vaporization near the cavity closure, resulting in strong spatially concentrated pressure-rate fluctuations. 
The close spatial correspondence between regions of strong mass transfer and elevated $dP/dt$ highlights the impulsive nature of cloud collapse events and their role in generating localized pressure transients. 

In contrast, at $T_2$, the wake-scale cavity is more elongated and the downstream
mass-transfer field is comparatively weaker. However, the near-wall
region highlighted by the purple box exhibits a sharply intensified
pressure-rate response associated with the inception and early growth
of a new cavity.
The acoustic response associated with this localized pressure-rate
amplification is next examined.


\begin{figure}[ht!]
	\centering
	\centering\includegraphics[width=0.8\linewidth]{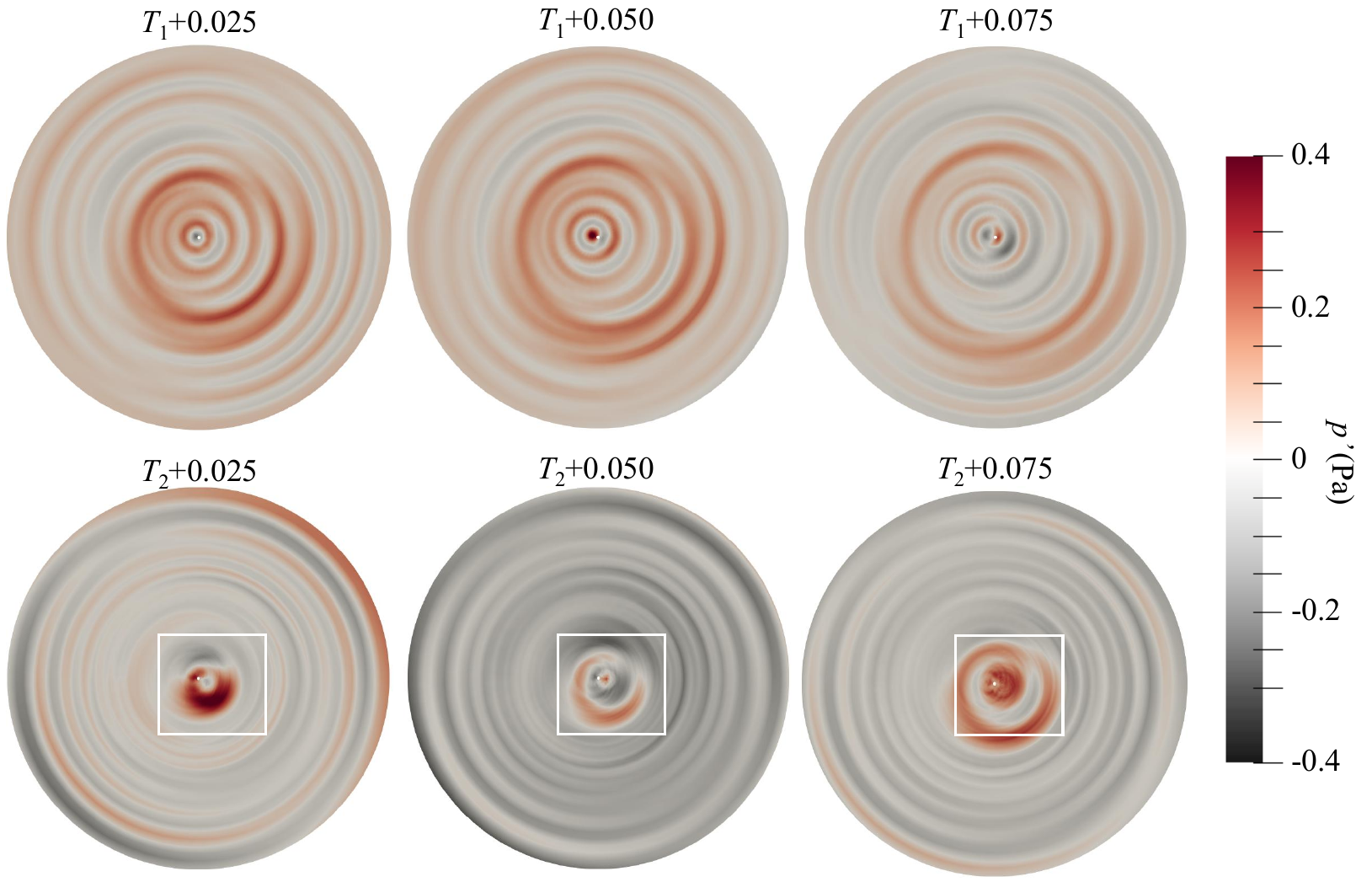}
    \caption{Instantaneous acoustic pressure perturbation fields $p'$ at representative phases $T_1$ and $T_2$ and subsequent snapshots, illustrating the emergence of localized tonal acoustic radiation.}
	\label{cylinder_Pp_contour}
\end{figure}

Figure~\ref{cylinder_Pp_contour} shows the instantaneous acoustic pressure perturbation field $p'$ in two representative phases, $T_1$ and $T_2$, together with three subsequent snapshots ($+0.025$, $+0.050$ and $+0.075$ in the same normalized time units). At $T_1$, the radiated field is relatively weak and appears as nearly concentric wavefronts with modest amplitude, consistent with predominantly broadband emission associated with downstream cloud shedding and collapse. 
In contrast, for non-cavitating flow passing a circular cylinder in Fig. \ref{cylinder_re200_COM}, the noise sources are surrounding the wake vortex-street, which is different from the cavitating situation, where the noise source is in the vicinity of the cylinder surface.

In contrast, at $T_2$ a pronounced spatially localized source region forms immediately adjacent to the cylinder surface (white box), and the subsequent snapshots reveal coherent wavefronts propagating outward with significantly enhanced amplitude. 
This marks the onset of a strong tonal component (`singing') and indicates that the dominant acoustic forcing is generated near the body surface rather than solely in the far wake.

Importantly, the emergence of this near-wall tonal source at $T_2$ is consistent with the preceding flow-field diagnostics: the previous figure showed that cavity inception/growth near the surface produces sharply elevated pressure-rate fluctuations ($dP/dt$) and intense local mass transfer, both of which act as efficient acoustic drivers. Within the CAPE framework for cavitating flows, the rapid phase change introduces volumetric source terms (monopole-like contributions through compressibility and dilatation associated with mass transfer), while unsteady loading and wall pressure fluctuations contribute to dipole-like forcing. The strong localized $p'$ generation observed in $T_2$ therefore reflects a coupled contribution in which the cavitation dynamics near the wall amplify local pressure transients and excite a coherent acoustic mode, leading to sustained tonal radiation. 
This provides supporting evidence linking the `singing' behavior to the surface-adjacent cavitation process identified above.

\begin{figure}[h]
	\centering
	\centering\includegraphics[width=1.0\linewidth]{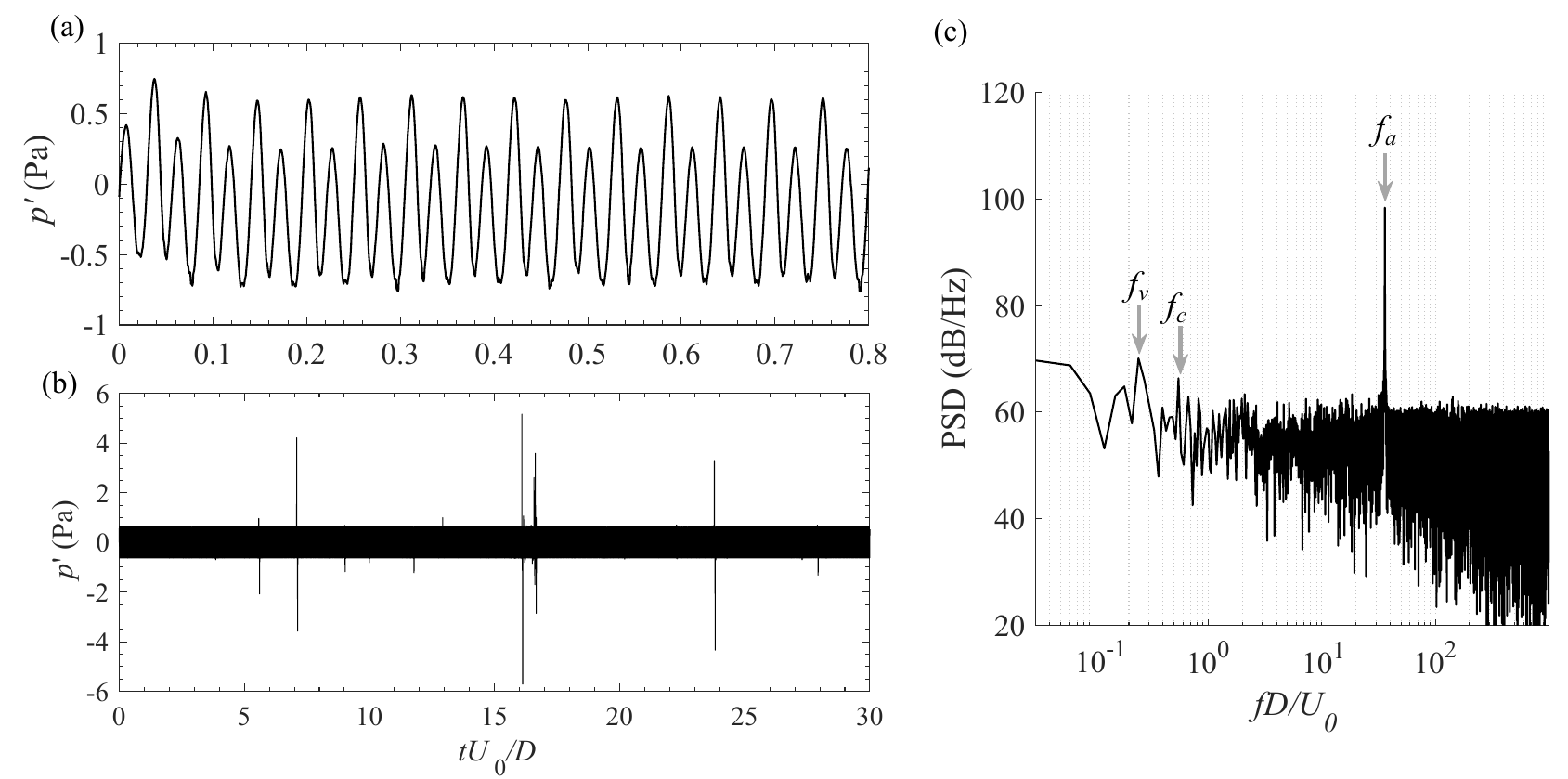}
	\caption{(a,b) Time histories in two scales and (c) spectra of acoustic pressure perturbation $p'$ at (0, 20$D$), highlighting intermittent impulsive events and the associated dominant tonal frequencies.}
	\label{Pp_time_PSD}
\end{figure}

Figure~\ref{Pp_time_PSD} further quantifies the acoustic response using the time history and spectrum of the acoustic pressure perturbation $p'$ at location (0, 20$D$). Panel~a shows a short-time segment in which $p'$ exhibits a nearly periodic oscillation, consistent with stable noise propagation at $T_1$ once the `singing' state is dissipated. When viewed over a longer duration (panel~(b)), the signal becomes strongly intermittent: ranges of sustained oscillations are interrupted by isolated, large-amplitude pressure impulses that recur quasi-periodically. These impulsive events coincide with the onset of the near-wall acoustic source observed at $T_2$ in Fig. \ref{cylinder_Pp_contour} and are consistent with the localized cavitation inception/growth process near the cylinder surface, which was shown to produce sharply elevated $dP/dt$ and intense mass transfer. Therefore, the long-time signal indicates that the singing is not continuously present, but is repeatedly triggered by surface-adjacent cavitation dynamics that inject strong pressure transients into the acoustic field.
Panel~c provides a clearer spectral separation of the relevant frequencies by marking the vortex-shedding frequency $f_v$, the cavitation shedding frequency $f_c$, and the acoustic eigenfrequency $f_a$ of the system. The most prominent feature is that the tonal peaks in the acoustic-pressure spectrum align with $f_c$ and its harmonics (e.g., $2f_c$, $3f_c$, \emph{etc.}), demonstrating that the `singing' behavior is fundamentally phase-locked to the cavitation shedding/collapse cycle rather than to the hydrodynamic vortex-shedding mode. $f_a$ is the specific frequency of spatial patterns at which sound waves naturally resonate within the given physical space or medium, not directly from vortex shedding or cavitation shedding.

\begin{figure}
	\centering
	\centering\includegraphics[width=0.8\linewidth]{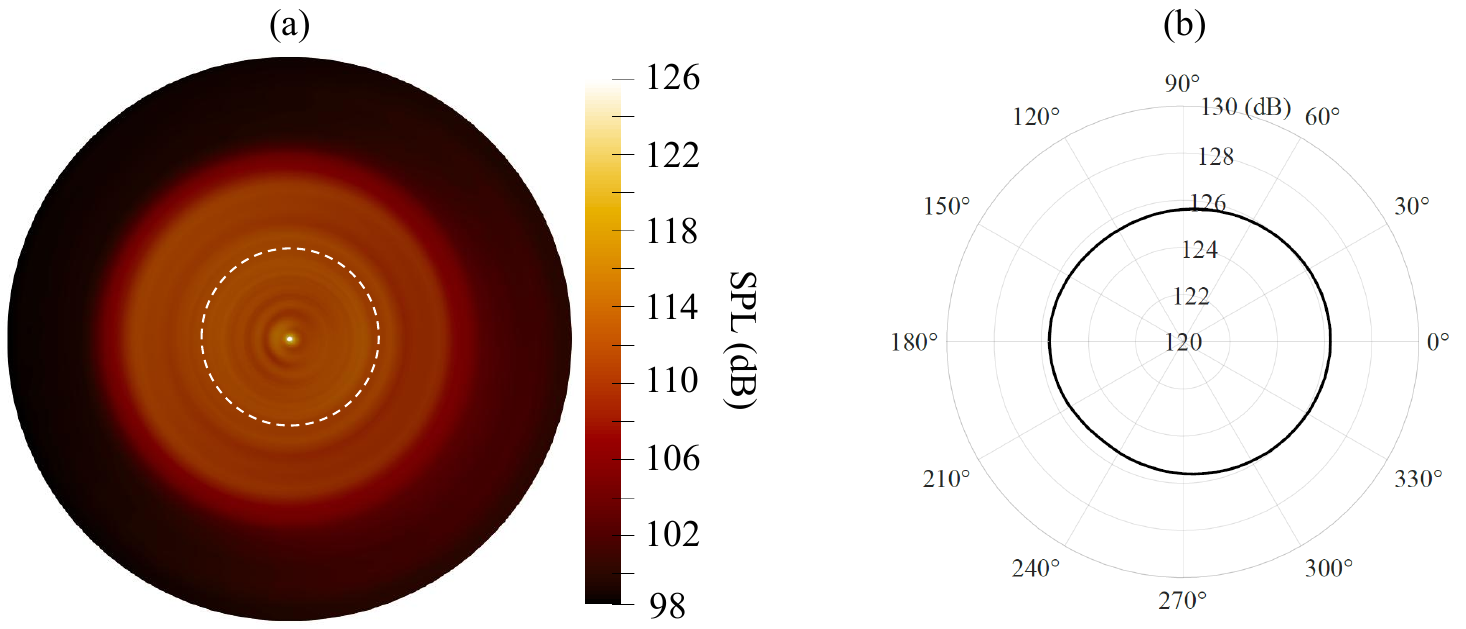}
	\caption{(a) Spatial distribution of sound pressure level ($SPL$) and (b) corresponding directivity pattern at a radius of 20$D$ (marked by white-dotted lines in panel (a)) for the cavitating circular cylinder, illustrating nearly axisymmetric radiation.}

	\label{SPL_contour_directicity}
\end{figure}

Figure~\ref{SPL_contour_directicity} presents the spatial distribution of $SPL$ (with \(P_{ref}\) as the underwater acoustic reference pressure of \(1\times 10^{-6} Pa\)) and the corresponding directivity pattern at a radius of 20$D$ (marked by a dotted line). As shown in Fig.~\ref{SPL_contour_directicity}a, the contour $SPL$ exhibits nearly concentric rings with weak angular dependence, indicating that the acoustic energy is radiated almost uniformly in all directions. This feature is further confirmed by the directivity plot in Fig.~\ref{SPL_contour_directicity}b, where the $SPL$ remains nearly constant throughout the azimuthal range, a characteristic signature of monopole-dominated radiation. This behavior is consistent with cavitation-induced acoustic sources, in which rapid volume changes associated with vapor generation and collapse act as effective monopole sources within the acoustic perturbation framework.

In contrast, the non-cavitating case exhibits a markedly different directivity pattern, as illustrated in Fig.~\ref{cylinder_re200_COM}. The $SPL$ shows a pronounced angular variation with a distinct lobe structure, closely showing a dipole-like radiation pattern that is characteristic of unsteady hydrodynamic loading and vortex shedding in bluff-body flows. The comparison highlights a fundamental shift in the dominant noise-generation response: while non-cavitating flow noise is primarily governed by dipole sources associated with fluctuating lift forces, the onset of cavitation introduces strong volumetric source terms that overwhelm the dipole contribution and lead to monopole-dominated acoustic radiation. This further supports the proposed cavitation-induced singing mechanism.

\begin{figure}[h]
	\centering
	\centering\includegraphics[width=0.8\linewidth]{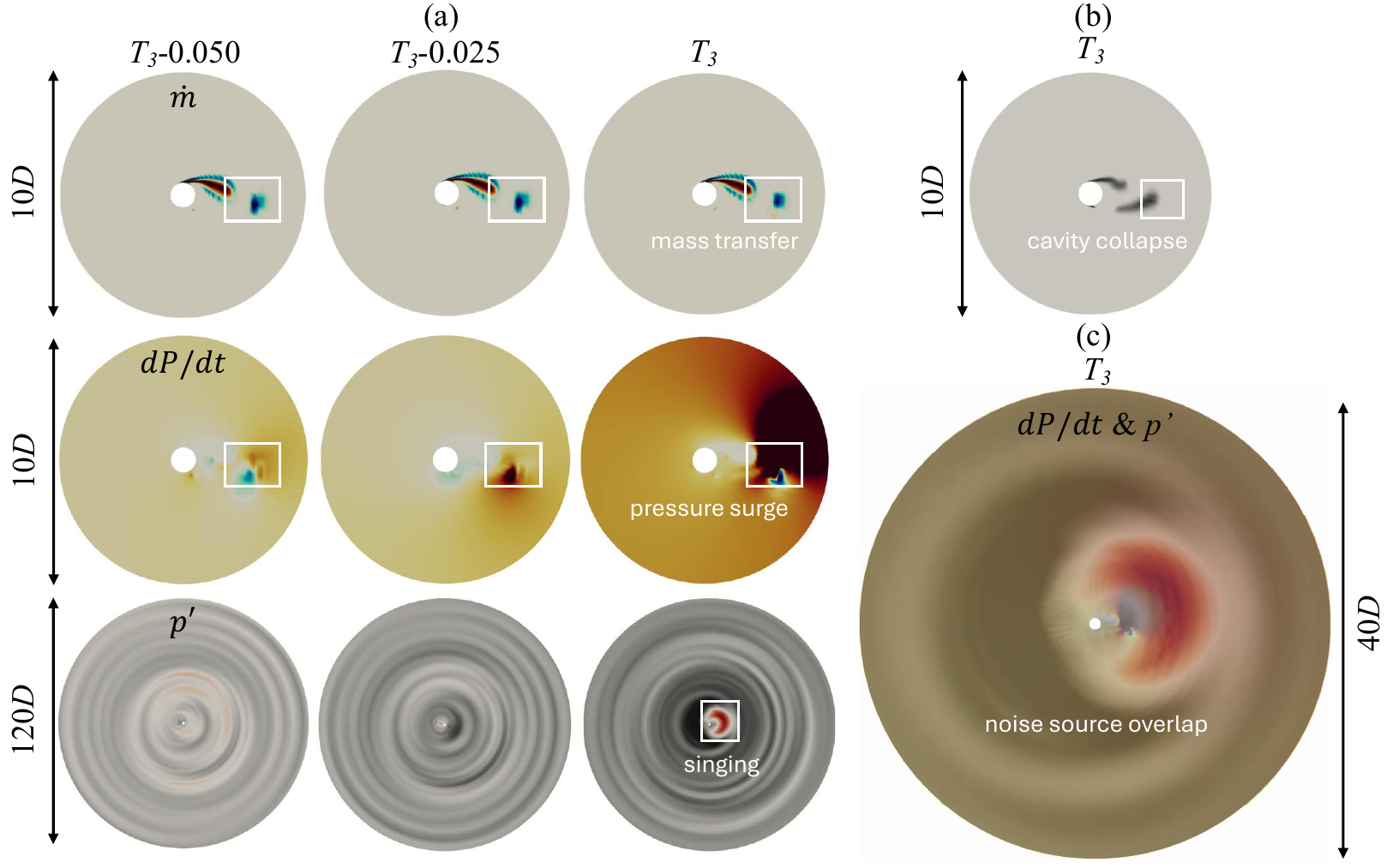}
	\caption{Cavitation collapse and acoustic source contribution at the end of the cavitation cycle. 
(a) Instantaneous contours of cavitation mass-transfer rate $\dot{m}$ (top), pressure time derivative $\mathrm{d}P/\mathrm{d}t$ (middle), and acoustic pressure perturbation $p'$ (bottom) at three successive instants around $T_3$, illustrating rapid mass transfer, pressure surge, and the onset of tonal acoustic radiation (singing). 
(b) Instantaneous cavity collapse pattern at $T_3$, highlighting the localized collapse region. 
(c) Combined visualization of $\mathrm{d}P/\mathrm{d}t$ and $p'$ at $T_3$, demonstrating spatial overlap between cavitation-induced pressure transients and acoustic source regions.}
	\label{collapse_and_singing}
\end{figure}

To further elucidate the generation of tonal noise, Fig.~\ref{collapse_and_singing} examines the flow and acoustic fields in the immediate temporal vicinity of a representative singing event that occurs at $T_3$. Fig.~\ref{collapse_and_singing}a presents a sequence of snapshots showing the cavitation mass-transfer rate $\dot{m}$, the local pressure time derivative $dP/dt$, and the acoustic pressure perturbation $p'$ at the instants preceding and including $T_3$. At $T_3-0.050$ and $T_3-0.025$, the contour of $\dot{m}$ remains largely intact, which means that there is no distinct collapse; correspondingly, $dP/dt$ remains relatively moderate, and the acoustic field exhibits only weak, broadband fluctuations. In contrast, at $T_3$, a sudden collapse of the cavity occurs within the highlighted region, accompanied by an abrupt intensification of mass transfer and a sharp surge in $dP/dt$, indicating a transient rapid pressure.

Fig.~\ref{collapse_and_singing}b isolates the cavity-collapse region at $T_3$, confirming that the collapse is spatially localized, yet dynamically violent, leading to a sudden release of compressive energy. This event directly triggers a strong acoustic response, as evidenced by the emergence of a high-amplitude localized acoustic pressure source. Fig.~\ref{collapse_and_singing}c further demonstrates the close spatial overlap between the elevated regions $dP/dt$ and the acoustic pressure $p'$, providing quantitative correspondence that collapse-induced pressure-rate amplification acts as the dominant contribution in the present source formulation. 
The resulting acoustic emission exhibits an intensity far exceeding that of the background sound waves. The singing frequency is consistent with the tonal peak observed at the cavitation shedding frequency. Together, these observations confirm that singing-like tonal features originate from discrete, near-wall cavity-collapse events that generate intense pressure transients, which are efficiently converted into coherent tonal noise through the coupled cavitation-acoustic correlation identified in the above analysis.

\begin{figure}[h]
	\centering
	\centering\includegraphics[width=1.0\linewidth]{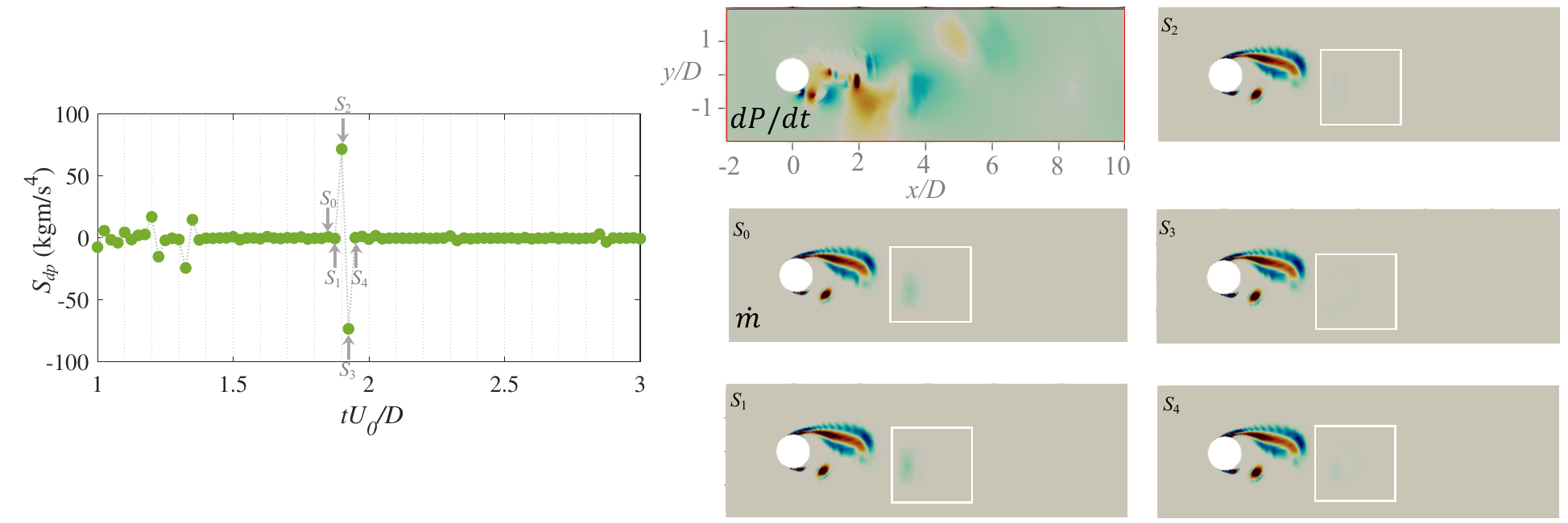}
	\caption{Temporal evolution of the pressure-rate source strength within the selected collapse region. 
Left: time history of $S_{dp}=\int_{\Omega_c}\partial P/\partial t\, d\Omega$. 
Right: corresponding contours of cavitation mass-transfer rate $\dot{m}$ at representative instants $S_0$-$S_4$. The red box denotes the integration region $\Omega_c$.}
	\label{singing_noise_counter}
\end{figure}

Figure~\ref{singing_noise_counter} provides direct evidence linking cavity collapse to the pressure-rate amplification responsible for the singing phenomenon. The quantity
\[
S_{dp}(t)=\int_{\Omega_c}\frac{\partial P}{\partial t}\,d\Omega
\]
is evaluated within the red-boxed region shown on the right. As illustrated by the contours of the vapor mass-transfer, the cavity structure remains visible at instants $S_0$ and $S_1$ inside the white box, while rapidly collapsing at $S_2$, indicating a localized collapse event. This instant coincides with the largest excursion of $S_{dp}$, followed by a peak of opposite sign at $S_3$. The strong temporal correspondence between cavity collapses and pressure-rate amplification suggests that rapid vapor-volume variation acts as the primary trigger for the acoustic source term. The results therefore support the interpretation that cavity collapse generates an intense source through the pressure-rate response identified in the present acoustic formulation, providing a source-term-level connection between localized cavitation dynamics and the computed tonal acoustic response.

\subsection{Cavitating Flow Passing Hydrofoil at High-Reynolds-Number Flow}
\label{}

\begin{figure}[h]
	\centering
	\centering\includegraphics[width=1.0\linewidth]{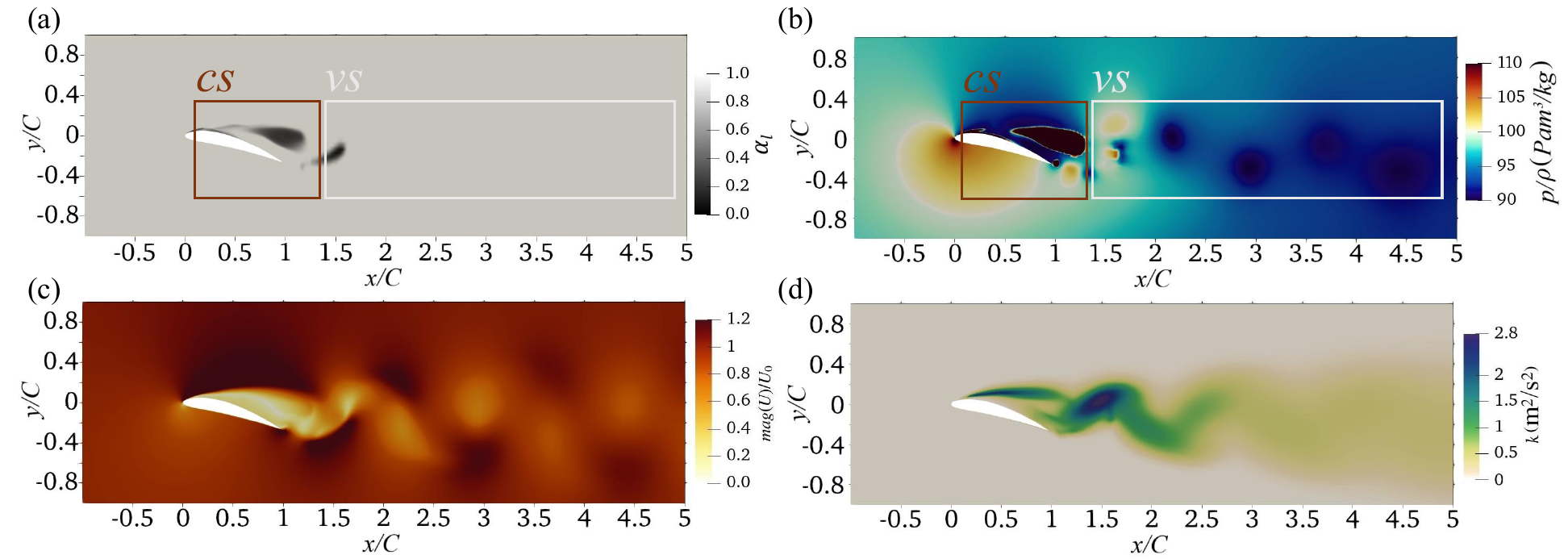}
	\caption{Instantaneous flow-field variables for cavitating flow past the hydrofoil. 
(a) Liquid phase fraction $\alpha_l$. 
(b) Pressure field normalized by fluid density, $p/\rho$. 
(c) Velocity magnitude, $mag(U)/U_0$. 
(d) Turbulent kinetic energy $k$.}
	\label{NACA_flow_fields}
\end{figure}

This subsection further considers a NACA 6412 hydrofoil with a chord length of 0.04 m and an angle of attack of 15$^\circ$; other parameters, including free-stream velocity, turbulence intensity, and cavitation number, are identical to those of the above-described configuration of cavitating flow past a cylinder.
Figure~\ref{NACA_flow_fields} presents instantaneous flow fields for cavitating flow past the hydrofoil, including (a) liquid volume fraction $\alpha_l$, (b) pressure normalized by density $p/\rho$, (c) normalized velocity magnitude $mag(U)/U_0$, and (d) turbulent kinetic energy $k$. Two distinct regions are highlighted: the cavitation-shedding region ($cs$) located near the suction side, and the downstream vortex-shedding region ($vs$) in the wake. Similarly to the cylinder case, cavitation shedding is characterized by the formation, detachment, and downstream convection of vapor structures, accompanied by strong local pressure gradients and rapid phase change near the solid boundary. In contrast, the vortex-shedding region exhibits alternating vortical structures and elevated turbulent kinetic energy, reflecting the hydrodynamic wake instability that primarily governs large-scale flow unsteadiness.

The spatial separation between $cs$ and $vs$ indicates a clear distinction between cavitation-dominated and vortex-dominated dynamics in the hydrofoil flow. While the $vs$ region contributes to wake unsteadiness and broadband fluctuations, the $cs$ region is associated with intense pressure variations and localized mass-transfer activity, suggesting a dominant role in impulsive loading and acoustic emission. This separation provides a useful framework for interpreting subsequent analyses of force spectra, cavitation dynamics, and noise generation for the hydrofoil.

\begin{figure}[h]
	\centering
	\centering\includegraphics[width=1.0\linewidth]{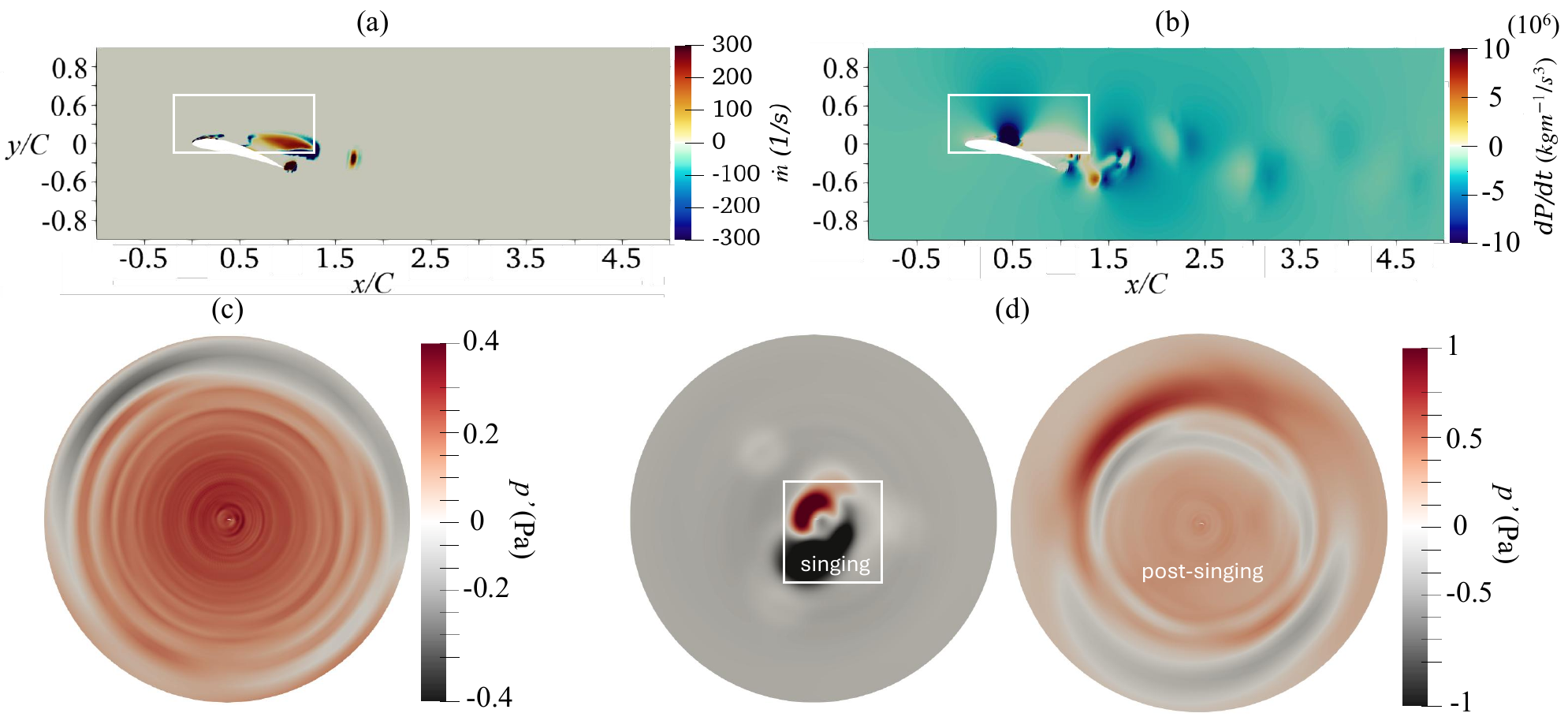}
	\caption{Coupled cavitation and acoustic source fields for the cavitating hydrofoil. 
(a) Instantaneous cavitation mass-transfer rate $\dot{m}$. 
(b) Instantaneous pressure time derivative $\mathrm{d}P/\mathrm{d}t$. 
(c) Acoustic pressure perturbation $p'$ at a representative time instant. 
(d) Comparison of acoustic pressure fields during the singing and post-singing states, illustrating changes in acoustic radiation patterns following the tonal emission.}
	\label{NACA_collapse_and_singing}
\end{figure}

Figure~\ref{NACA_collapse_and_singing}a shows the instantaneous cavitation mass-transfer rate, where intense vaporization and condensation are concentrated near the cavity closure on the suction side of the hydrofoil. These localized mass-transfer events accompany the onset of cavity collapse, highlighted by the white box in Fig. \ref{NACA_collapse_and_singing}b, which shows a sharp amplification of the local pressure time derivative $dP/dt$. The strong pressure-rate surge is analogous to the collapse-driven response identified in the circular-cylinder case.
The acoustic consequences of this event are illustrated in Fig. \ref{NACA_collapse_and_singing}c and d. Prior to collapse, the acoustic pressure field is dominated by relatively smooth, broadband radiation with weak spatial localization, as shown in Fig. \ref{NACA_collapse_and_singing}c. At the moment of collapse, however, a highly localized, high-amplitude acoustic source emerges near the hydrofoil surface (Fig. \ref{NACA_collapse_and_singing}d, `singing'), indicating the generation of a coherent tonal component. 
The computed tonal component is correlated with the collapse-induced pressure-rate amplification and is consistent with the cylinder case.
Following this event, the acoustic field transitions into a post-singing state characterized by outward-propagating wavefronts. 
These results show that the proposed formulation captures localized collapse-associated acoustic generation for the hydrofoil configuration as well.
However, since cavitation is generated on the suction side for a hydrofoil, in contrast to the cylinder case where cavity behavior is symmetrical, it is expected that the $SPL$ directivity for a hydrofoil will be spatially asymmetrical.

\begin{figure}[h]
	\centering
	\centering\includegraphics[width=0.8\linewidth]{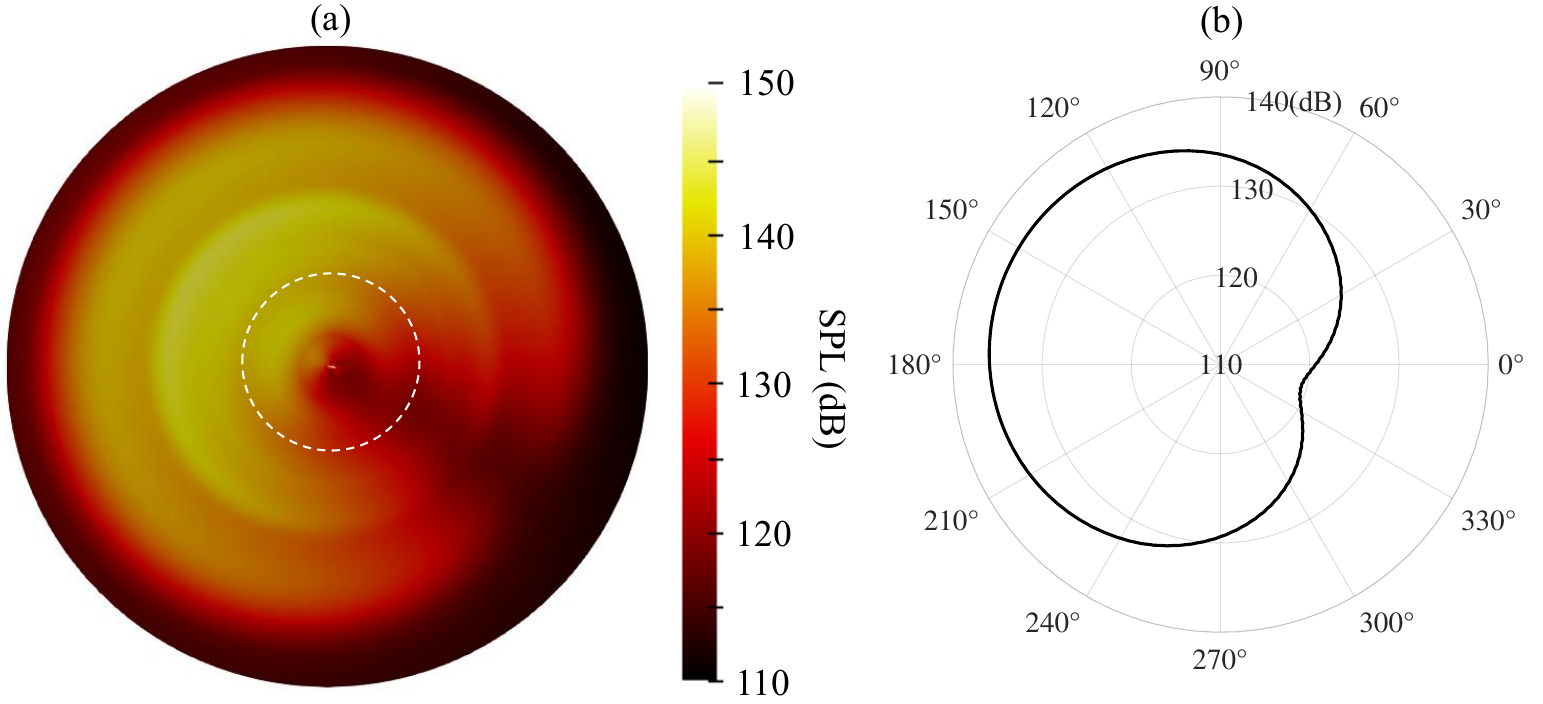}
	\caption{Sound pressure level ($SPL$) distribution for the cavitating hydrofoil. 
(a) $SPL$ field in the acoustic domain, illustrating the spatial distribution of radiated sound and the shielding effect of the hydrofoil geometry. 
(b) Far-field $SPL$ directivity pattern at radius of 10$C$, showing the angular dependence of acoustic radiation influenced by geometric blocking by the hydrofoil.}

	\label{NACA_SPL_contour_directicity}
\end{figure}

Figure~\ref{NACA_SPL_contour_directicity} presents the spatial distribution of $SPL$ and the corresponding directivity evaluated along the dotted-line circle marked on panel~(a). Unlike the nearly axisymmetric radiation observed for the cavitating circular cylinder, the hydrofoil case exhibits a distinctly asymmetric $SPL$ field. As shown in panel~(a), the highest levels of $SPL$ are concentrated near the suction side of the hydrofoil, where cavitation shedding and collapse occur predominantly. This asymmetry is further reflected in the directivity pattern shown in panel~(b), which displays a clear angular dependence with enhanced radiation toward the suction-side half-plane and attenuated levels in the opposite direction.

The asymmetric directivity arises from the combined effects of localized cavitation sources and geometric shielding by the hydrofoil surface. Since the dominant noise sources are situated near the suction side, the solid body partially blocks and reflects acoustic waves propagating toward the pressure side, reducing the effective radiation in that direction. Consequently, the acoustic field deviates from a purely monopole-like pattern and exhibits a skewed radiation characteristic shaped by both source localization and wall-induced shadowing. For a rotating propeller, blade rotation and temporal averaging would
be expected to reduce the fixed-frame asymmetry observed for the
stationary hydrofoil.

\section{Conclusion}
This work presents a cavitation-aware acoustic framework for hydroacoustic prediction in liquid-vapor multiphase flows. Based on homogeneous mixture equations, the new Cavitation Acoustic Perturbation Equations (CAPE) framework is derived by incorporating cavitation mass transfer and mixture compressibility into the perturbation system. The resulting CAPE preserves the structure of conventional APE solvers while introducing additional source terms associated with phase change, enabling cavitation-induced acoustic generation to be resolved within the computational domain. A corresponding finite-volume predictor-corrector algorithm with acoustic subcycling and perfectly matched layer treatment is developed for the numerical solution of the CAPE governing equations.

The numerical implementation is first verified through one-dimensional propagation problems. The convergence of the grid and the time-steps, frequency preservation, PML performance, and acoustic attenuation are systematically evaluated. The predicted attenuation follows Stokes' sound attenuation law over a range of source frequencies, demonstrating that the proposed solver accurately reproduces both wave propagation and dissipation characteristics while maintaining numerical stability.

The framework is subsequently applied to cavitating flow past a circular cylinder and a NACA hydrofoil. The simulations capture the evolution of cavitation dynamics together with the corresponding acoustic field and reproduce the characteristic tonal radiation observed in cavitating flows. Compared with conventional formulations, the proposed method directly incorporates cavitation-induced volumetric source terms into the governing equations, allowing hydrodynamic loading, phase-change dynamics, and acoustic propagation to be solved within a unified computational framework. 
The predicted acoustic fields further demonstrate that the method can distinguish geometry-dependent radiation patterns in bluff-body and lifting-body configurations while retaining the cavitation-induced source terms explicitly in the governing equations.

The present formulation provides a computationally efficient extension of APE-based hydroacoustic prediction to cavitating flows while remaining compatible with existing incompressible flow solvers. 
The methodology provides a practical numerical framework for hydroacoustic prediction, source-field analysis, and future extension to more complex cavitating-flow configurations. 
Future work will extend the formulation to three-dimensional turbulent cavitation, rotating propulsors, and fluid-structure interaction to investigate cavitation noise in realistic marine propulsion systems.

\section*{Acknowledgments}

The first author is supported by the Natural Sciences and Engineering Research Council of Canada (NSERC). This work was made possible by the facilities of Shared Hierarchical Academic Research Computing  (\href{http://www.sharcnet.ca}{SHARCNET}), and Compute/Calcul Canada.

\section*{Competing interests}
The authors declare no competing financial interests.

\appendix

\section{Face-Level Finite-Volume Discretization}
\label{app:face_discretization}
This appendix provides the control-volume integration and face-level
interpolation underlying the compact predictor-corrector system
presented in Section~\ref{sec_PC}.
Consider the acoustic co-velocity equation
\begin{equation}
\frac{\partial \boldsymbol{f}}{\partial t}
+
\nabla\cdot
\left[
\boldsymbol{f}
\left(
\boldsymbol{U}+\boldsymbol{u}'
\right)
+
\rho_0\boldsymbol{U}\boldsymbol{u}'
\right]
=
-\nabla p'
+
\nabla\cdot\boldsymbol{\tau}'.
\label{eq:appendix_momentum}
\end{equation}
Integrating the transient term over the control volume $V_P$ and the
time interval $[t,t+\Delta t]$ gives
\begin{equation}
\int_t^{t+\Delta t}
\int_{V_P}
\frac{\partial \boldsymbol{f}}{\partial t}
\,\mathrm{d}V\,\mathrm{d}t
=
\left(
\boldsymbol{f}_P^{*}
-
\boldsymbol{f}_P^{t}
\right)V_P .
\label{eq:appendix_transient}
\end{equation}
The convective contribution is discretized as
\begin{align}
&\int_t^{t+\Delta t}
\int_{V_P}
\nabla\cdot
\left[
\boldsymbol{f}
\left(
\boldsymbol{U}+\boldsymbol{u}'
\right)
+
\rho_0\boldsymbol{U}\boldsymbol{u}'
\right]
\,\mathrm{d}V\,\mathrm{d}t
\nonumber\\
&\qquad =
\Delta t
\sum_{f\in\partial V_P}
\left[
\boldsymbol{f}_f^{*}
\left(
\boldsymbol{U}_f^{t}
+
\boldsymbol{u}_f'^{\,t}
\right)
+
\rho_{0,f}^{t}
\left(
\boldsymbol{U}_f^{t}
\boldsymbol{u}_f'^{\,t}
\right)
\right]
\cdot\boldsymbol{S}_f .
\label{eq:appendix_convection}
\end{align}

The perturbed viscous-stress contribution becomes
\begin{equation}
\int_t^{t+\Delta t}
\int_{V_P}
\nabla\cdot\boldsymbol{\tau}'
\,\mathrm{d}V\,\mathrm{d}t
=
\Delta t
\sum_{f\in\partial V_P}
\boldsymbol{\tau}_f'^{\,*}
\cdot\boldsymbol{S}_f ,
\label{eq:appendix_viscous}
\end{equation}
while the pressure-gradient contribution is
\begin{equation}
\int_t^{t+\Delta t}
\int_{V_P}
\nabla p'
\,\mathrm{d}V\,\mathrm{d}t
=
\Delta t
\sum_{f\in\partial V_P}
p_f'^{\,t}\boldsymbol{S}_f .
\label{eq:appendix_pressure}
\end{equation}
Combining
Eqs.~\eqref{eq:appendix_transient}-\eqref{eq:appendix_pressure}
yields
\begin{align}
\frac{V_P}{\Delta t}
\left(
\boldsymbol{f}_P^{*}
-
\boldsymbol{f}_P^{t}
\right)
&+
\sum_{f\in\partial V_P}
\left[
\boldsymbol{f}_f^{*}
\left(
\boldsymbol{U}_f^{t}
+
\boldsymbol{u}_f'^{\,t}
\right)
+
\rho_{0,f}^{t}
\left(
\boldsymbol{U}_f^{t}
\boldsymbol{u}_f'^{\,t}
\right)
\right]
\cdot\boldsymbol{S}_f
\nonumber\\
&=
\sum_{f\in\partial V_P}
\boldsymbol{\tau}_f'^{\,*}
\cdot\boldsymbol{S}_f
-
\sum_{f\in\partial V_P}
p_f'^{\,t}\boldsymbol{S}_f .
\label{eq:appendix_discrete_balance}
\end{align}
Here, $\boldsymbol{S}_f$ is the outward-oriented face-area vector, the
subscript $f$ denotes a face-centered quantity, and the superscript
$*$ denotes the predicted acoustic field.
The face-centered co-velocity and pressure are obtained using
arithmetic interpolation:
\begin{equation}
\boldsymbol{f}_f^{*}
=
\frac{
\boldsymbol{f}_P^{*}
+
\boldsymbol{f}_N^{*}
}{2},
\label{eq:appendix_f_interpolation}
\end{equation}
and
\begin{equation}
p_f'^{\,t}
=
\frac{
p_P'^{\,t}
+
p_N'^{\,t}
}{2},
\label{eq:appendix_p_interpolation}
\end{equation}
where $N$ denotes the neighboring cell sharing face $f$ with cell $P$.

Substituting
Eqs.~\eqref{eq:appendix_f_interpolation} and
\eqref{eq:appendix_p_interpolation} into
Eq.~\eqref{eq:appendix_discrete_balance} gives
\begin{align}
\frac{V_P}{\Delta t}
\left(
\boldsymbol{f}_P^{*}
-
\boldsymbol{f}_P^{t}
\right)
&+
\sum_{f\in\partial V_P}
\left[
\frac{
\boldsymbol{f}_P^{*}
+
\boldsymbol{f}_N^{*}
}{2}
\left(
\boldsymbol{U}_f^{t}
+
\boldsymbol{u}_f'^{\,t}
\right)
+
\rho_{0,f}^{t}
\left(
\boldsymbol{U}_f^{t}
\boldsymbol{u}_f'^{\,t}
\right)
\right]
\cdot\boldsymbol{S}_f
\nonumber\\
&=
\sum_{f\in\partial V_P}
\boldsymbol{\tau}_f'^{\,*}
\cdot\boldsymbol{S}_f
-
\sum_{f\in\partial V_P}
\frac{
p_P'^{\,t}
+
p_N'^{\,t}
}{2}
\boldsymbol{S}_f .
\label{eq:appendix_interpolated_balance}
\end{align}
Collecting the coefficients multiplying the cell-centered co-velocity
unknowns produces
\begin{equation}
A_P^{t}\boldsymbol{f}_P^{*}
+
\sum_{N\in\mathcal{N}(P)}
A_N^{t}\boldsymbol{f}_N^{*}
-
\boldsymbol{E}_P^{t}
=
-\left(\nabla p'^{\,t}\right)_P ,
\label{eq:appendix_algebraic_predictor}
\end{equation}
which is Eq.~\eqref{eq:co_velocity_predictor} in the main text.
During the correction step, the corrected face co-velocity is
reconstructed as
\begin{equation}
\boldsymbol{f}_f^{**}
=
\mathrm{\mathcal{H}}_f^{*}
-
\frac{1}{A_f^{t}}
\left(\nabla_f p'^{\,*}\right),
\label{eq:appendix_face_correction}
\end{equation}
where the face-centered quantities
$\mathrm{\mathcal{H}}_f^{*}$ and $1/A_f^{t}$ are obtained by interpolation
from the neighboring cell-centered values. The corrected face fluxes
are then used to enforce the discrete acoustic continuity equation.

\bibliography{mybibfile}

\end{document}